\documentclass[epj]{svjour}
\usepackage{epsfig}
\usepackage{graphicx}

\def\rootfig{./}

\begin{document}

\title{Dynamics of trapped bright solitons in the presence of localized
inhomogeneities}
\author{ G. Herring\inst{1}, P. G.\ Kevrekidis\inst{1}, R.
Carretero-Gonz\'{a}lez\inst{2},
B. A. Malomed\inst{3}, D. J. Frantzeskakis\inst{4}, and A. R.
Bishop\inst{5} } \institute{\inst{1} Department of Mathematics and
Statistics, University of
Massachusetts, Amherst MA 01003-4515, USA \\
\inst{2} Nonlinear Dynamical Systems
Group\thanks{http://nlds.sdsu.edu/}, Department of Mathematic and
Statistics, San Diego
State University, San Diego CA, 92182-7720,\\
\inst{3} Department of Interdisciplinary Studies, Tel Aviv University, Tel
Aviv 69978, Israel \\
\inst{4} Department of Physics, University of Athens, Panepistimiopolis,
Zografos, Athens 15784, Greece \\
\inst{5} Center for Nonlinear Studies and Theoretical Division, Los Alamos
National Laboratory, Los Alamos, NM 87545, USA\\
}
                                                                               
%
\date{Submitted to {\em Eur.\ Phys.\ J.\ D: At.\ Mol.\ Opt.\ Phys.}, March 2005\\
This version has low quality graphs.\\
For high quality go to: http://www.rohan.sdsu.edu/$\sim$rcarrete/ [Publications] [Publication{\#}41]}
%
\abstract{ We
examine the dynamics of a bright solitary wave in the presence of
a repulsive or attractive localized ``impurity'' in Bose-Einstein
condensates (BECs). We study the generation and stability of a
pair of steady states in the vicinity of the impurity as the
 impurity strength is varied. These two new steady
states, one stable and one unstable, disappear through a
saddle-node bifurcation as the strength of the impurity is
decreased. The dynamics of the soliton is also examined in all the
cases (including cases where the soliton is offset from  one of
the relevant fixed points). The numerical results are corroborated
by theoretical calculations which are in very good agreement with
the numerical findings.
}
\titlerunning{Dynamics of trapped bright solitons in the presence of
localized inhomogeneities} \authorrunning{G.\ Herring {\it et al.}}

\maketitle

\section{Introduction\label{sec:intro}}

In the past few years, the rapid experimental and theoretical developments 
in the field of Bose-Einstein condensates (BECs) \cite{reviews} have led to a 
surge of interest in the study of the nonlinear matter waves 
that appear in this context. More specifically, experiments have yielded bright 
solitons in self-attractive condensates ($^{7}$Li) in a nearly 
one-dimensional setting \cite{expb}, as well as their dark \cite{expd} and,
more recently, gap \cite{Oberthaler} counterparts in repulsive condensates, 
such as $^{87}$Rb. The study of these matter-wave solitons, apart from being
a topic of interest in its own right, may also have important applications.
For instance, a soliton may be transferred and manipulated similarly to what
has been recently shown, experimentally and theoretically, for BECs in
magnetic waveguides \cite{aw} and atom chips \cite{ac}. Furthermore,
similarities between matter and light waves suggest that some of the
technology developed for optical solitons \cite{ka} may be adjusted for
manipulations with MWs, and thus applied to the rapidly evolving field of
quantum atom optics (see, e.g., \cite{mo}).

One of the topics of interest in this context is how 
matter-waves can be steered/manipulated by means of external potentials, 
currently
available experimentally. In addition to the commonly known
magnetic trapping of the atoms in a parabolic potential, it is
also experimentally feasible to have a sharply focused laser beam,
such as ones already used to engineer desired density
distributions of BECs in experiments \cite{expd}. Depending on
whether it is blue-detuned or red-detuned, this beam repels or
attracts atoms, thus generating a localized ``defect" that can
induce various types of the interaction with solitary matter waves. 
This possibility was developed to some extent in theoretical
\cite{theor} and experimental \cite{exper} studies of dynamical
effects produced by moving defects, such as the generation of gray
solitons and sound waves in one dimension \cite{rad2325}, and
formation of vortices in two dimensions (see, e.g., \cite{mplb} and references therein). 
The interaction of dark solitons with a localized 
impurity was also studied \cite{dimitri}.

Our aim here is to examine the interaction of a bright 
matter-wave soliton
with a strongly localized (in fact, $\delta $-like) defect, in the
presence of the magnetic trap. Our approach is different from that
of Ref. \cite{dimitri}, in that we will view the presence of the
defect as a bifurcation problem. We demonstrate that the localized
perturbation (independently of whether it is attractive or
repulsive) creates an effective potential that results in two
additional localized states (one of which is naturally stable,
while the other is always unstable) for sufficiently large
impurity strength. As one may expect on grounds of the general
bifurcation theory, these states will disappear, ``annihilating"
with each other, as the strength of the impurity is decreased
below a threshold value. We will describe this \textit{saddle-node
bifurcation} in the present context. We will also compare our
numerical predictions for its occurrence with analytical results
following from an approximation that treats the soliton as a
quasi-particle moving an effective potential. Very good
agreement between the analytical and numerical results will be
demonstrated. Finally, we will examine the dynamics of solitons
inside the combined 
potential, jointly created by the
magnetic trap and the localized defect. Both equilibrium positions and
motion of the free soliton will be considered in the latter case.

The paper is structured as follows: in Section \ref{sec:theo}, we present
our effective potential theory. In Section \ref{sec:num}, we discuss
numerical methods and results, and provide their comparison with the
analytical predictions. Finally, in Section \ref{sec:conc}, we summarize our
findings and present our conclusions.

\section{Setup and Theoretical Results\label{sec:theo}}

In the mean-field approximation, the single-atom wavefunction for a dilute
gas of ultra-cold atoms very accurately obeys the Gross-Pitaevskii equation
(GPE). Although the GPE naturally arises in the three-dimensional (3D)
settings, it has been shown \cite{theory,Isaac,Salasnich} that it can be
reduced to its one-dimensional (1D) counterpart for the so-called
cigar-shaped condensates. Cigar-shaped condensates are created when two
transverse directions of the atomic cloud are tightly confined, and the
condensate is effectively rendered one-dimensional, by suppressing dynamics
in the transverse directions. The effective equation describing this
quasi-1D is simply tantamount to a directly written 1D GPE. In normalized
units, it takes the well-known from,
\begin{equation}
iu_{t}=-\frac{1}{2} u_{xx}+g|u|^{2}u+V(x)u,  
\label{meq1}
\end{equation}
where subscripts denote partial derivatives and  
$u(x,t)$ is the one-dimensional mean-field wave function.
The normalized 
1D atomic density 
is given by $n=|u(x,t)|^{2}$, while the total number of atoms 
is proportional to the norm of the normalized wavefunction $u(x,t)$, which is an 
integral of motion of Eq. (\ref{meq1}): 
\begin{equation} 
P=\int_{-\infty }^{+\infty}|u(x,t)|^{2}dx.  
\label{N}
\end{equation}
The nonlinear coefficient in Eq. (\ref{meq1}) is $g = \pm 1$, 
for repulsive or attractive interatomic interactions respectively.
Finally, the magnetic trap, together with the localized defect, are described by a 
combined potential $V(x)$ of the form
\begin{equation}
V(x)=\frac{1}{2} \Omega^{2} x^{2}-V_{0}\delta (x-\xi ).  
\label{V}
\end{equation}

In Eqs.\ (\ref{meq1}) and (\ref{V}), the space variable $x$ is 
given in units of the healing length $\tilde{\xi}=\hbar /\sqrt{n_{0}g_{1D}m}$, 
where $n_{0}$ is the peak density, and the
normalized atomic density is measured in units of $n_{0}$. Here, the nonlinear 
coefficient is considered to have an effectively 1D form, namely 
$g_{1D}\equiv g_{3D}/(2\pi l_{\perp }^{2})$, 
where $g_{3D}=4\pi \hbar ^{2}a/m$ is the original 3D interaction strength
($a$ is the scattering length, $m$ is
the atomic mass, and $l_{\perp }=\sqrt{\hbar /m\omega _{\perp }}$
is the transverse harmonic-oscillator length, with $\omega _{\perp}$ 
being the transverse-confinement frequency). Further, time $t$
is given in units of $\tilde{\xi}/c$ (where
$c=\sqrt{n_{0}g_{1D}/m}$ is the Bogoliubov speed of sound), and
the energy is measured in units of the chemical potential, $\mu =g_{1D}n_{0}$. 
Accordingly, the dimensionless
parameter $\Omega \equiv \hbar \omega _{x}/g_{1D}n_{0}$ (where
$\omega _{x}$ is the confining frequency in the axial direction) 
determines the effective strength of the magnetic trap in the 1D
rescaled equations. Positive and negative values of $V_{0}$
corresponds, respectively, to the attractive and repulsive defect.
Finally, since we are interested in bright matter-wave solitons, which exist in the case of attraction, 
we hereafter set the normalized nonlinear coefficient $g=-1$.

It is worth mentioning that modified versions of the 1D GPE are known too.
One of them features a non-polynomial nonlinearity, instead of the cubic one
in Eq. (\ref{meq1}) \cite{Salasnich}. A different equation was derived for a
case of a very strong nonlinearity, so that the local value of the potential
energy exceeds the transverse kinetic energy. It amounts to the same cubic
equation (\ref{meq1}), but with a noncanonical normalization condition, with
the integral in Eq. (\ref{N}) replaced by $\int_{-\infty }^{+\infty}\left\vert u(x,t)\right\vert ^{4}dx$.

In the absence of potential, Eq.\ (\ref{meq1}) supports stationary soliton
solutions of the form
\begin{equation}
u_{s}(x)=\eta \,\mathrm{sech} \left[ \eta (x-\zeta ) \right] \exp \left( i\eta
^{2}t/2\right)   
\label{meq3}
\end{equation}
where $\eta $ is an arbitrary amplitude and $\zeta $ is the position of the
soliton's center. It is possible to generate moving solitons (with constant
velocity) by application of the Galilean boost to the stationary soliton in
Eq.\ (\ref{meq3}).

One can examine the persistence and dynamics of the bright solitary waves in the
presence of the potential $V(x)$ by means of the standard perturbation
theory (see e.g., Ref. \cite{RMP,sb} and a more rigorous approach, based on
the Lyapunov-Schmidt reduction, that was developed in Ref. \cite{kapitula}).
This method, which treats the soliton as a particle, yields effective
potential forces acting on the particle from the defect and the magnetic trap,
\begin{equation}
F_{\mathrm{def}}=2\eta ^{3}\tanh (\eta (\xi -\zeta
))\,\mathrm{sech}^{2}(\eta (\xi -\zeta ))V_{0},  \label{meq6}
\end{equation}\begin{equation}
F_{\mathrm{trap}}=-2\Omega ^{2}\zeta \eta ,  \label{meq7}
\end{equation}which enter the equation of motion for $\zeta (t)$:
\begin{equation}
\ddot{\zeta}=F_{\mathrm{imp}}+F_{\mathrm{trap}}.  \label{meq8}
\end{equation}Below, results following from this equation will be compared to direct
simulations of Eq.\ (\ref{meq1}).

The stationary version of Eq. (\ref{meq8}) ($\ddot{\zeta}=0$),
\begin{equation}
\left( \eta ^{3}V_{0}/\Omega ^{2}\right) \left( \tanh \theta
\right) \mathrm{sech}^{2}\theta =\eta \xi -\theta ,  \label{meq9}
\end{equation}with $\theta \equiv \eta (\xi -\zeta )$, determines equilibrium
positions ($\zeta $) of the soliton's center. Depending on parameters, this equation may
have one or three physical solutions, see below. In what follows, we will
examine the solutions in detail and compare them to numerical results
stemming from direct simulations of Eq. (\ref{meq1}).

\section{Numerical Methods and Results\label{sec:num}}

In order to numerically identify standing wave 
solutions of Eq.\ (\ref{meq1}), we
substitute, as usual, $u(x,t)=\exp (i\Lambda t)w(x)$, which results in the
steady-state problem:
\begin{equation}
\Lambda w=\frac{1}{2} w_{xx}+w^{3}-V(x)w.  \label{meq4a}
\end{equation}This equation is solved by a fixed-point iterative scheme on a
fine finite-difference grid. Then, we analyze the stability of the obtained
solutions by using the following ansatz for the perturbation,
\begin{equation}
u(x)=e^{i\Lambda t}\left[ w(x)+a(x)e^{-\lambda t}+b^{\ast }(x)e^{-\lambda
^{\star }t}\right]   \label{meq10}
\end{equation}(the asterisk stands for the complex conjugation), and solving the resulting
linearized equations for the perturbation eigenmodes $\{a(x),b(x)\}$ and
eigenvalues $\lambda $ associated with them. The resulting solutions are
also used to construct initial conditions for direct numerical simulations
of Eq.\ (\ref{meq1}), to examine typical scenarios of the full dynamical
evolution. To eliminate effects of the radiation backscattering in these
simulations, we have used absorbing boundary layers, by adding a term in Eq.\ (\ref{meq1}), of the
form:
\begin{equation}
\begin{array}{rcl}
\Gamma u & = & -\,\left[ (1+\tanh (1000(x-R))\right.  \\[2ex]
&  & \left. +\,\,\,(1-\tanh (1000(x-L))\right] \,u, \end{array}
\label{BCEqn}
\end{equation}
which is defined on the domain $L<x<R$. In the numerical
simulations presented herein, the $\delta $-function of the potential was
approximated by a Gaussian waveform, according to the well-known formula,
\begin{equation}
\delta (x)=\lim_{\sigma \rightarrow 0^{+}}\frac{1}{\sqrt{2\pi }\sigma }\exp
\left( -\frac{x^{2}}{4\sigma }\right) .  \label{delta}
\end{equation}
Lorentzian and hyperbolic-function approximations to the $\delta $-function
were also used, without producing any conspicuous difference in the results.

\begin{figure}[tbp]
\begin{center}
\includegraphics[width=4.3cm,height=3.5cm,angle=0,clip]{\rootfig
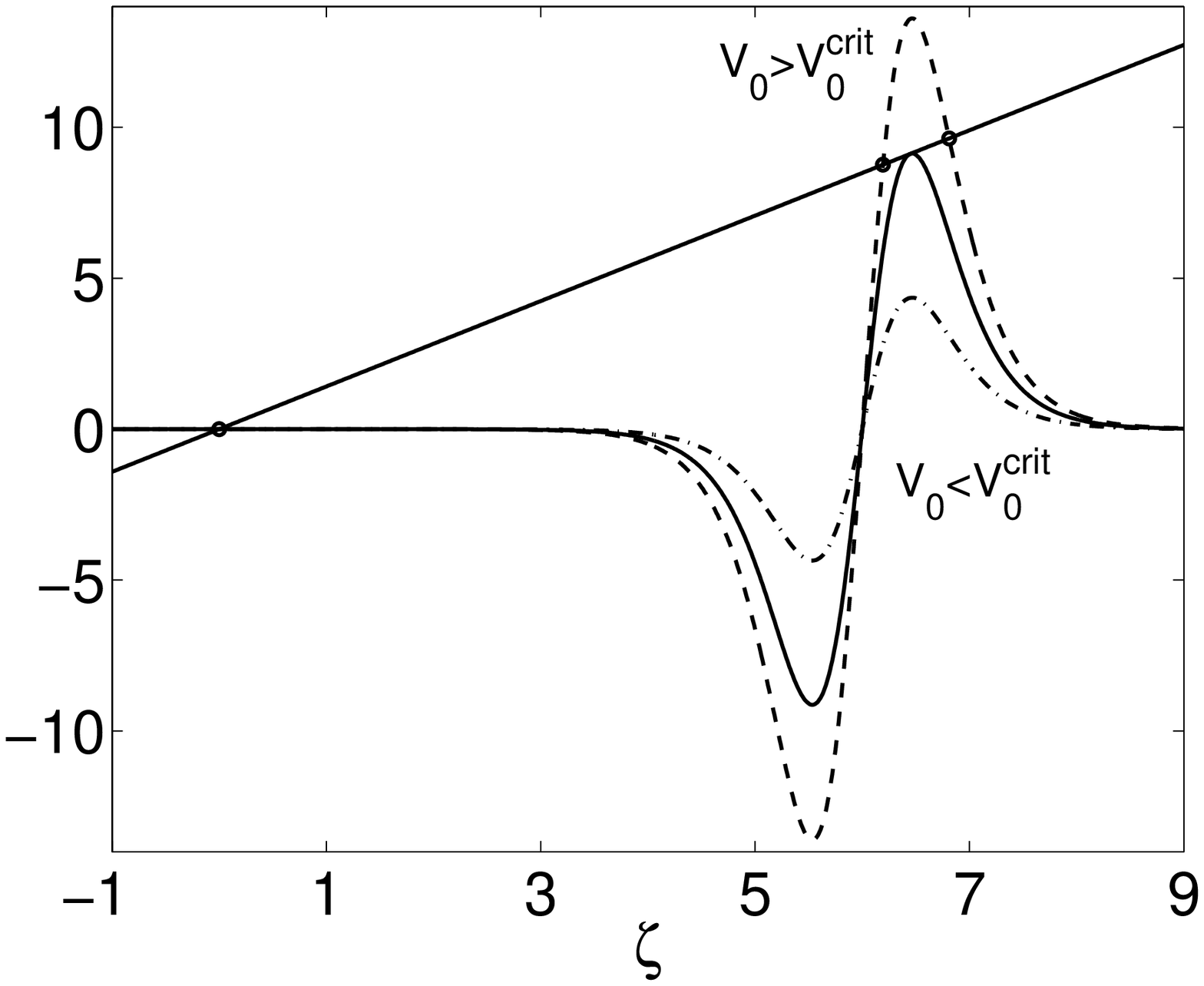} \includegraphics[width=4.3cm,height=3.6cm,angle=0,clip]{\rootfig 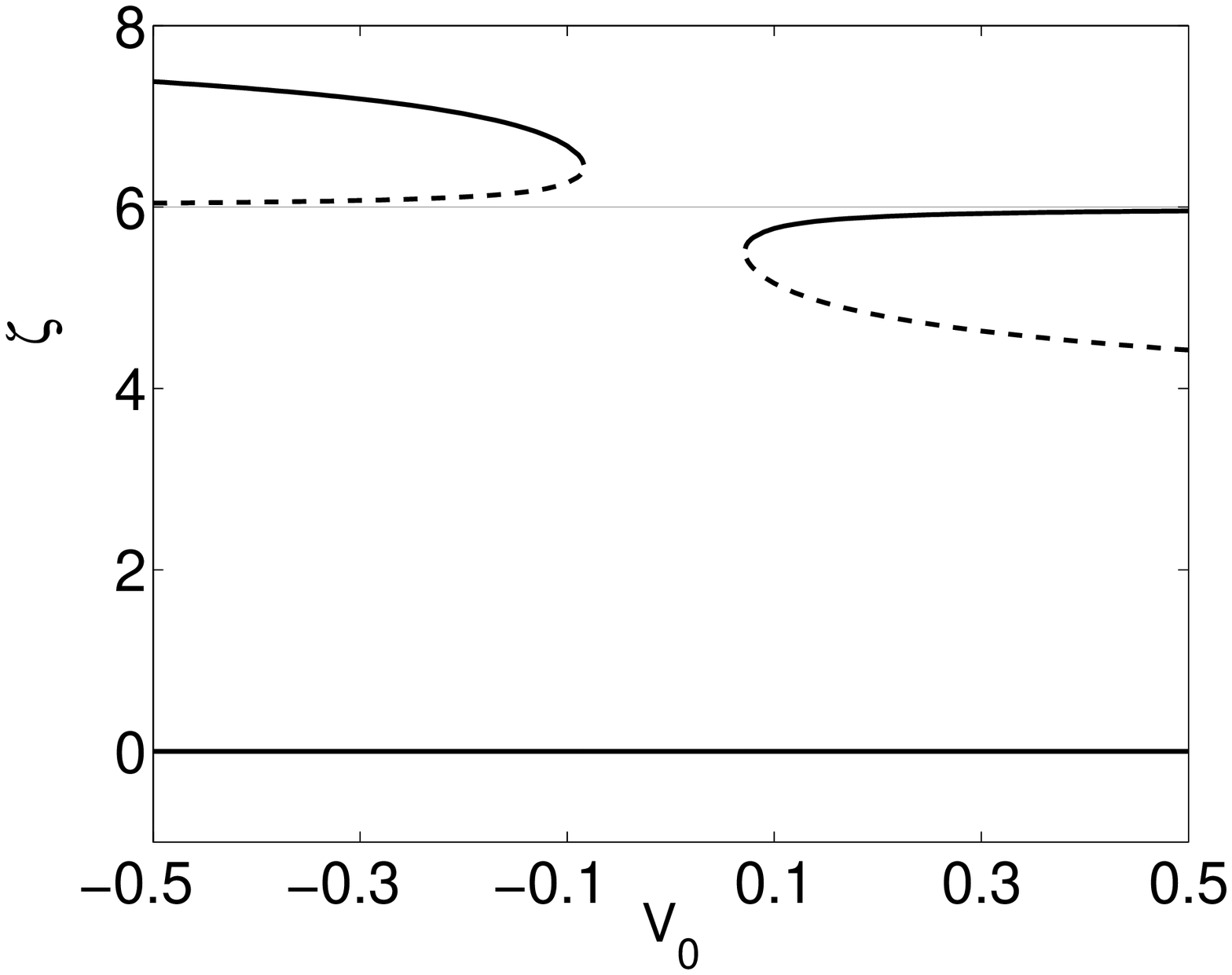} \\[1ex]
\includegraphics[width=4.3cm,height=3.5cm,angle=0,clip]{\rootfig 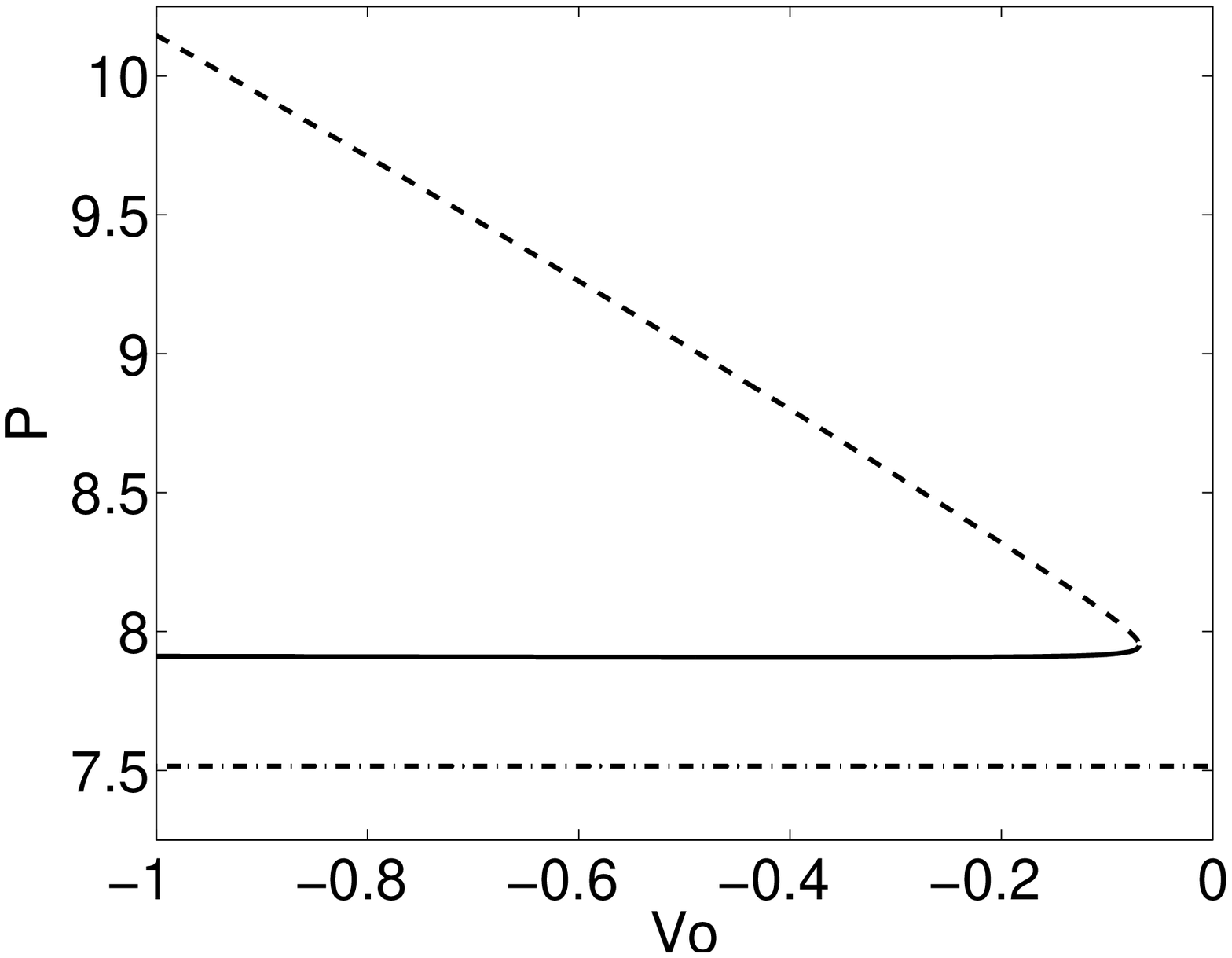}
\includegraphics[width=4.3cm,height=3.6cm,angle=0,clip]{\rootfig 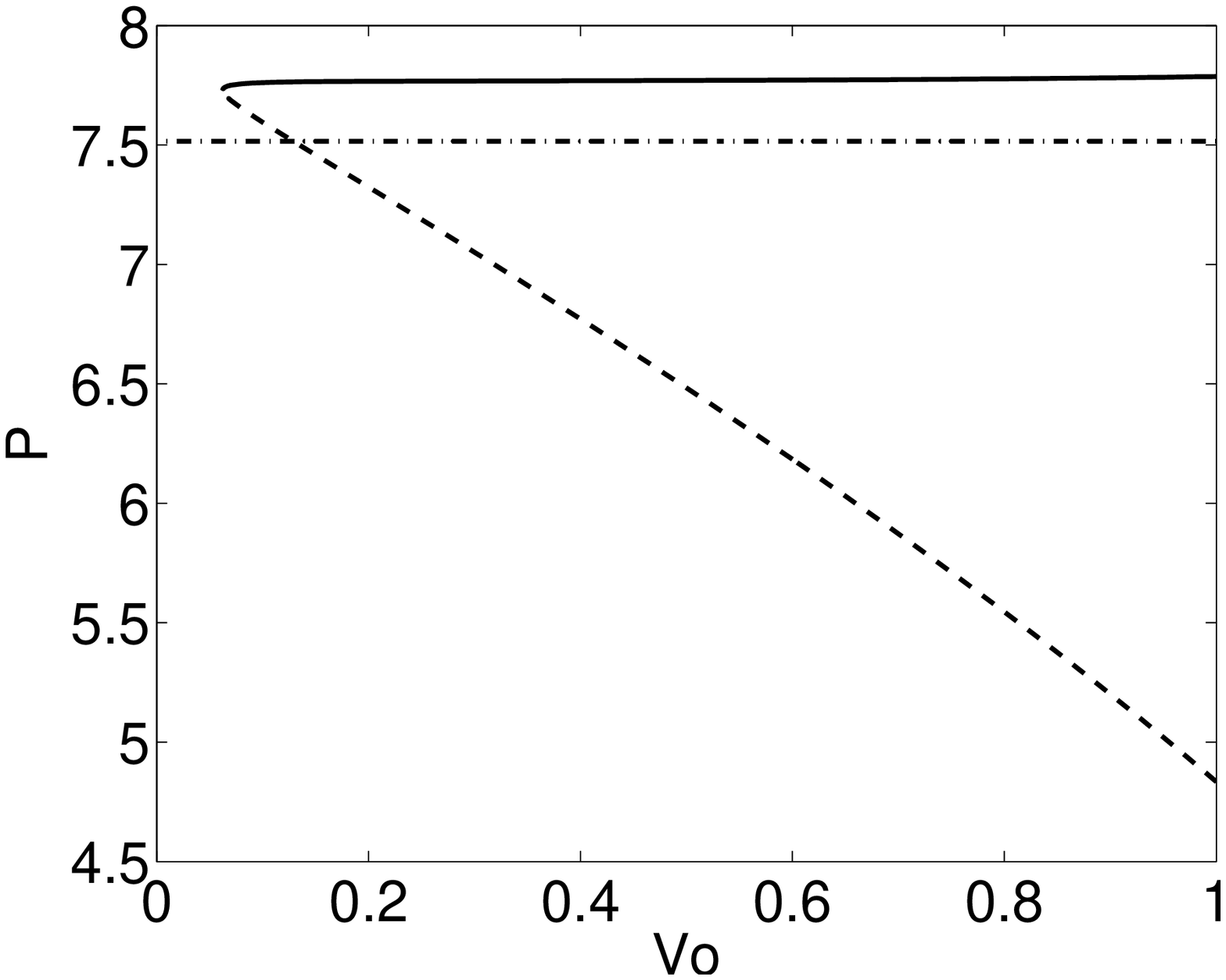}
\end{center}
\caption{Saddle-node bifurcation of stationary states for the
location $\protect\zeta $ of the bright soliton inside the
magnetic trap ($\Omega =0.1$), with the localized defect of
strength $V_{0}$ located at $\protect\xi =6$. The top-left panel
displays the corresponding solutions of the stationary equation\
(\protect\ref{meq9}). For the weak defect, dash-dotted line, only
one steady state exists very close to the origin. As the defect's
strength increases, two additional fixed points (both located on
the same side of the impurity) are created in a saddle-node
bifurcation. The top-right panel depicts the position and
stability (solid for stable and dashed for unstable) for the
steady states as a function of the defect's strength $V_{0} $. The
thin horizontal line for $\protect\xi =6$ shows the location of
the defect. The two bottom plots depict another version of the
stability diagram, in terms of the soliton's norm $P$ [see Eq.
(\protect\ref{N})], as $V_{0}$ is varied.} \label{FigSaddleNode}
\end{figure}

As mentioned above, depending on the value of the defect's strength $V_{0}$,
Eq.\ (\ref{meq9}) may have either one or three physical roots for $\zeta $
(the equilibrium position of the soliton's center). The border between these
two generic cases is a separatrix where two of the roots merge in one before
they disappear. All the qualitatively different cases are depicted in the
top-left panel of Fig.\ \ref{FigSaddleNode}. 
The physical interpretation of this result can be given as follows.
Obviously, in the absence of the defect there exists a stable solitary-wave
configuration centered at $\zeta =0$ (hence there is a single steady state
in the problem). On the other hand, it is easy to see that Eq.\ (\ref{meq9})
generates three solutions for large $V_{0}$. Hence, there should be a
bifurcation point, of the saddle-node type, that leads to the disappearance
of two branches of the solutions as $V_{0}$ decreases. Furthermore, based on
general bifurcation theory principles, one of the steady states
disappearing as a result of the bifurcation may correspond to a stable
soliton, whereas its companion branch definitely represents an unstable
solitary wave. The full bifurcation-diagram scenario, for the position of
the soliton's center and its norm, is depicted in Fig.\ \ref{FigSaddleNode}.
It is interesting to note that the norm of the solitons corresponding to the
unstable branches varies almost linearly with the defect's strength, while
the stable branches correspond to solitons whose norm is approximately
constant.

\begin{figure}[t]
\begin{center}
\includegraphics[width=5.cm,height=4cm,angle=0,clip]{\rootfig 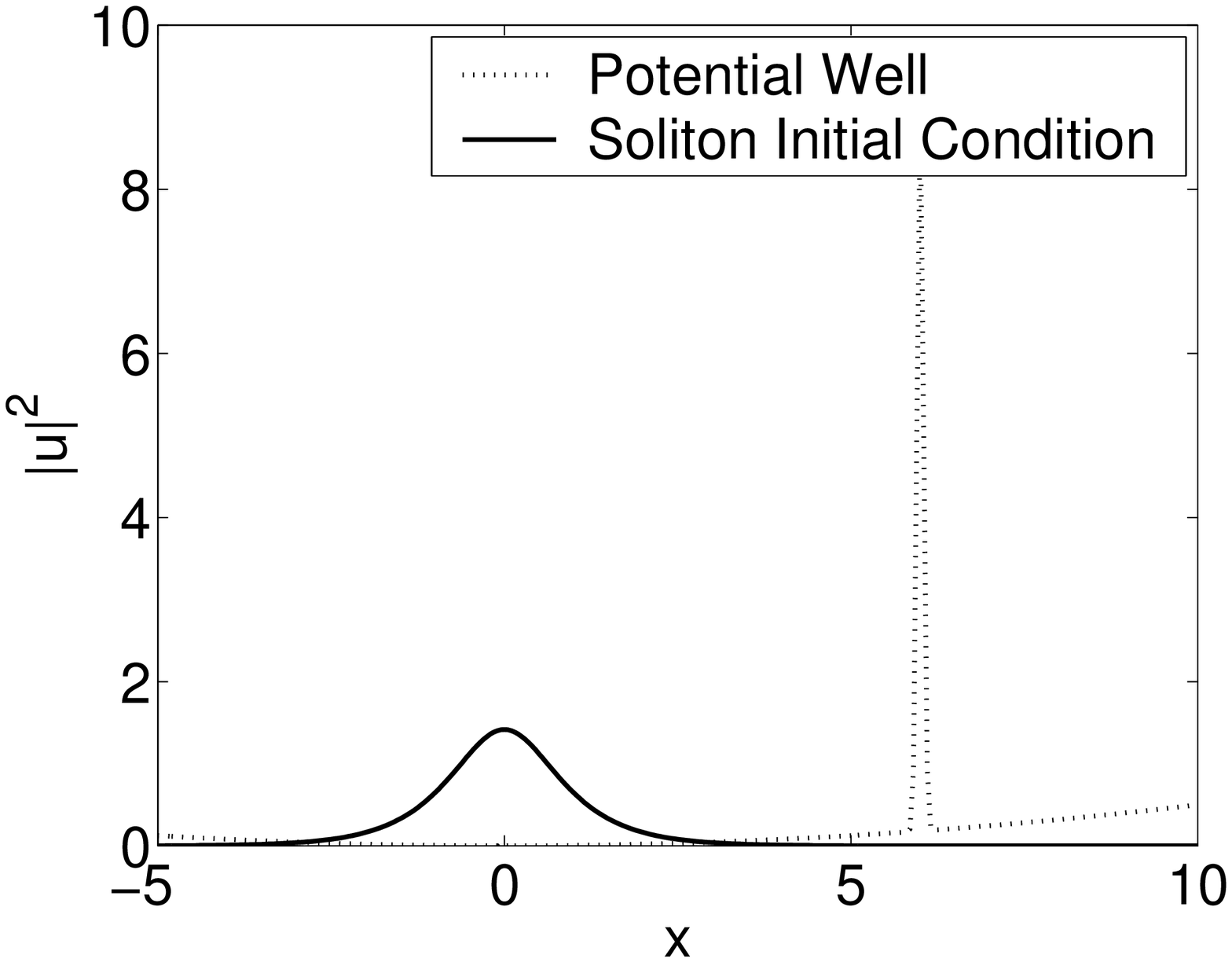}
\includegraphics[width=5.cm,height=4cm,angle=0,clip]{\rootfig
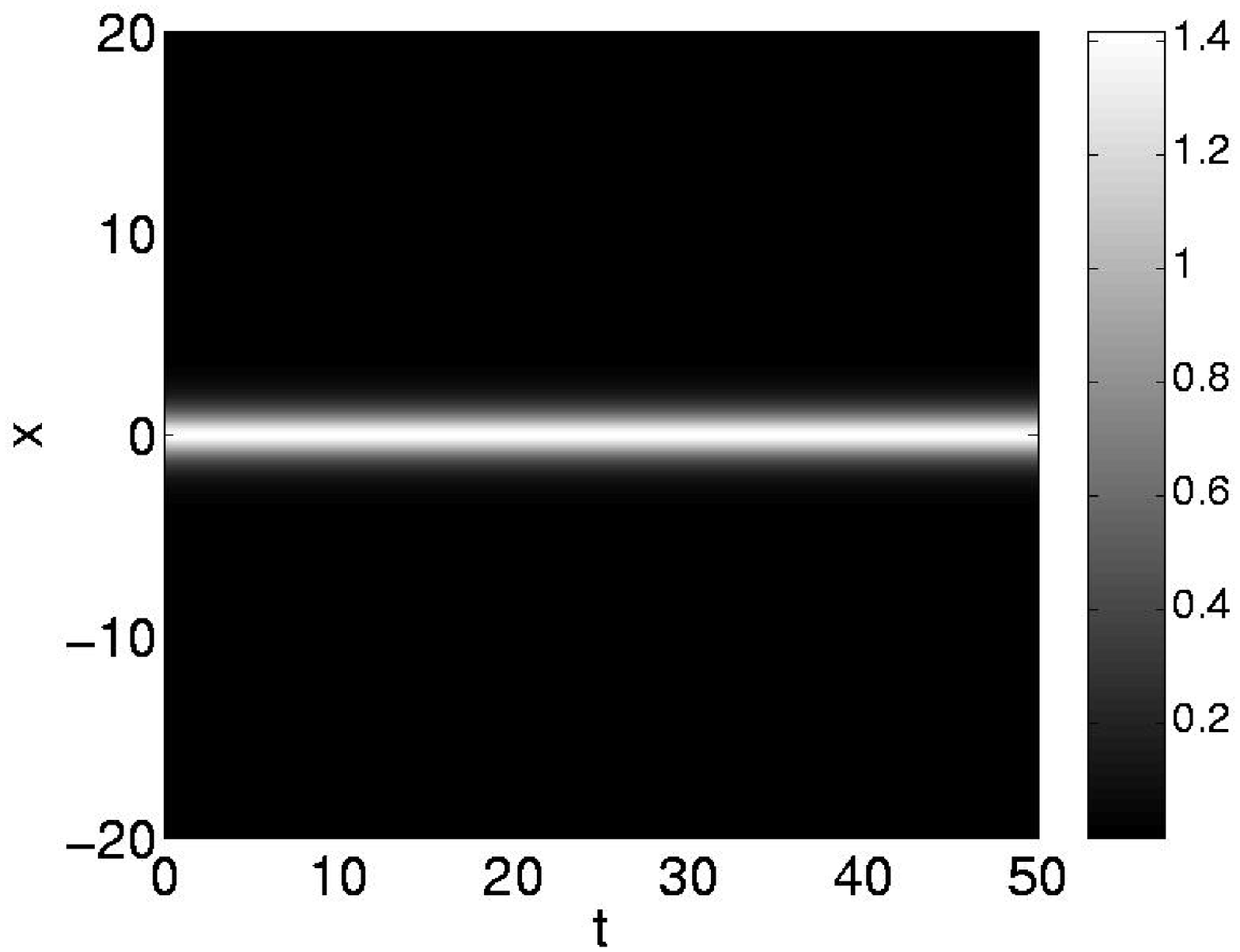} \includegraphics[width=5.cm,height=4cm,angle=0,clip]{\rootfig 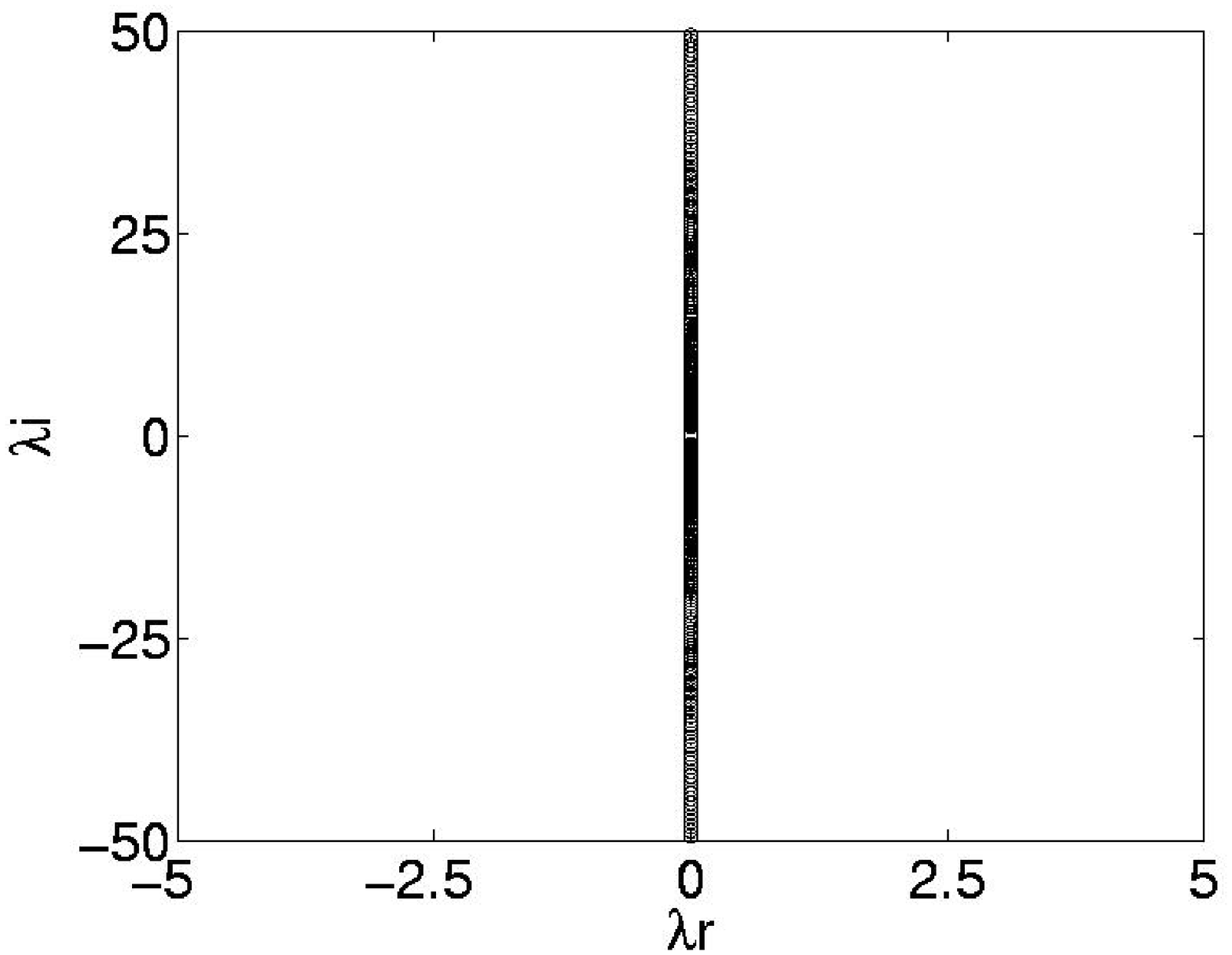} \\[0pt]
\includegraphics[width=5.cm,height=4cm,angle=0,clip]{\rootfig 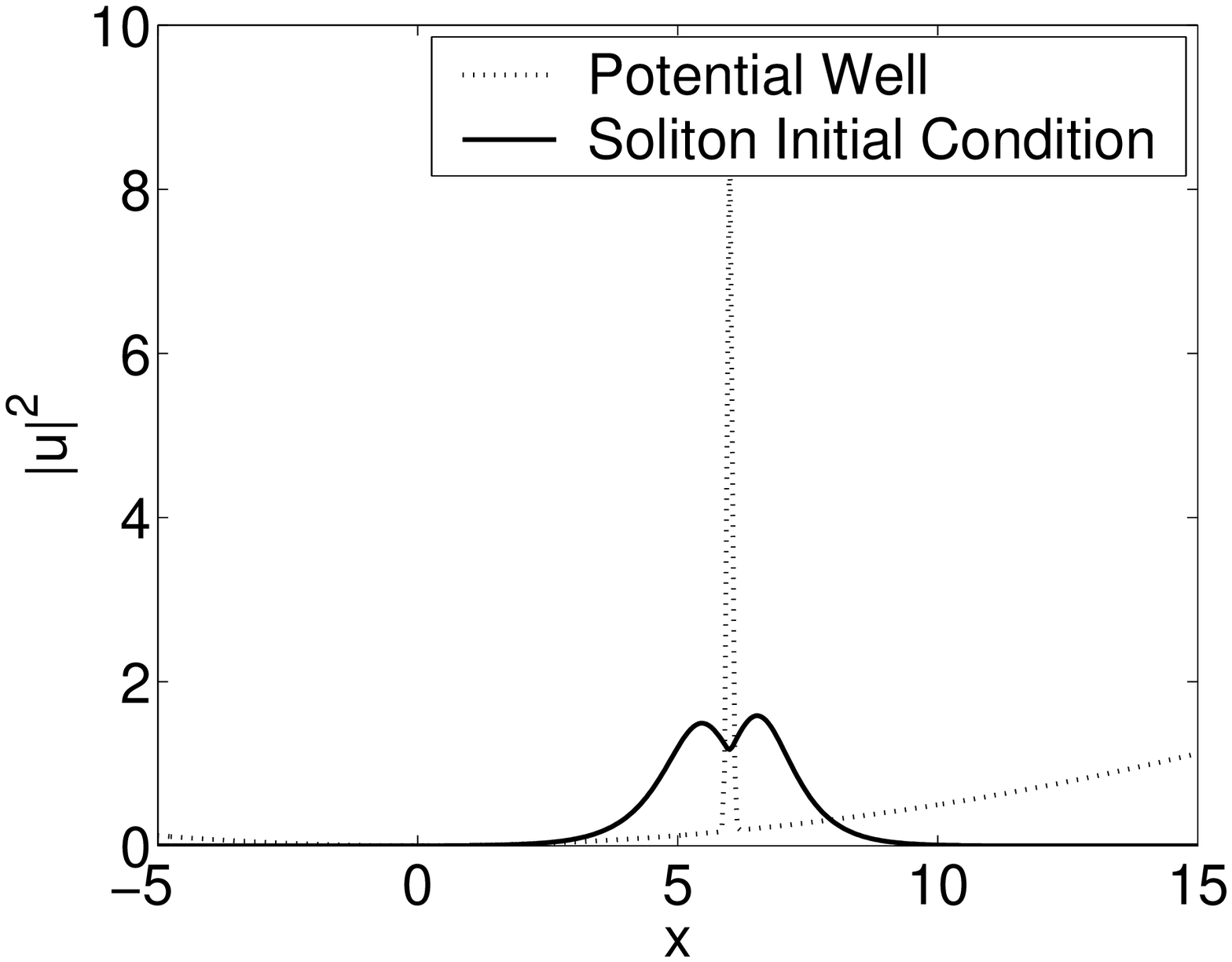}
\includegraphics[width=5.cm,height=4cm,angle=0,clip]{\rootfig
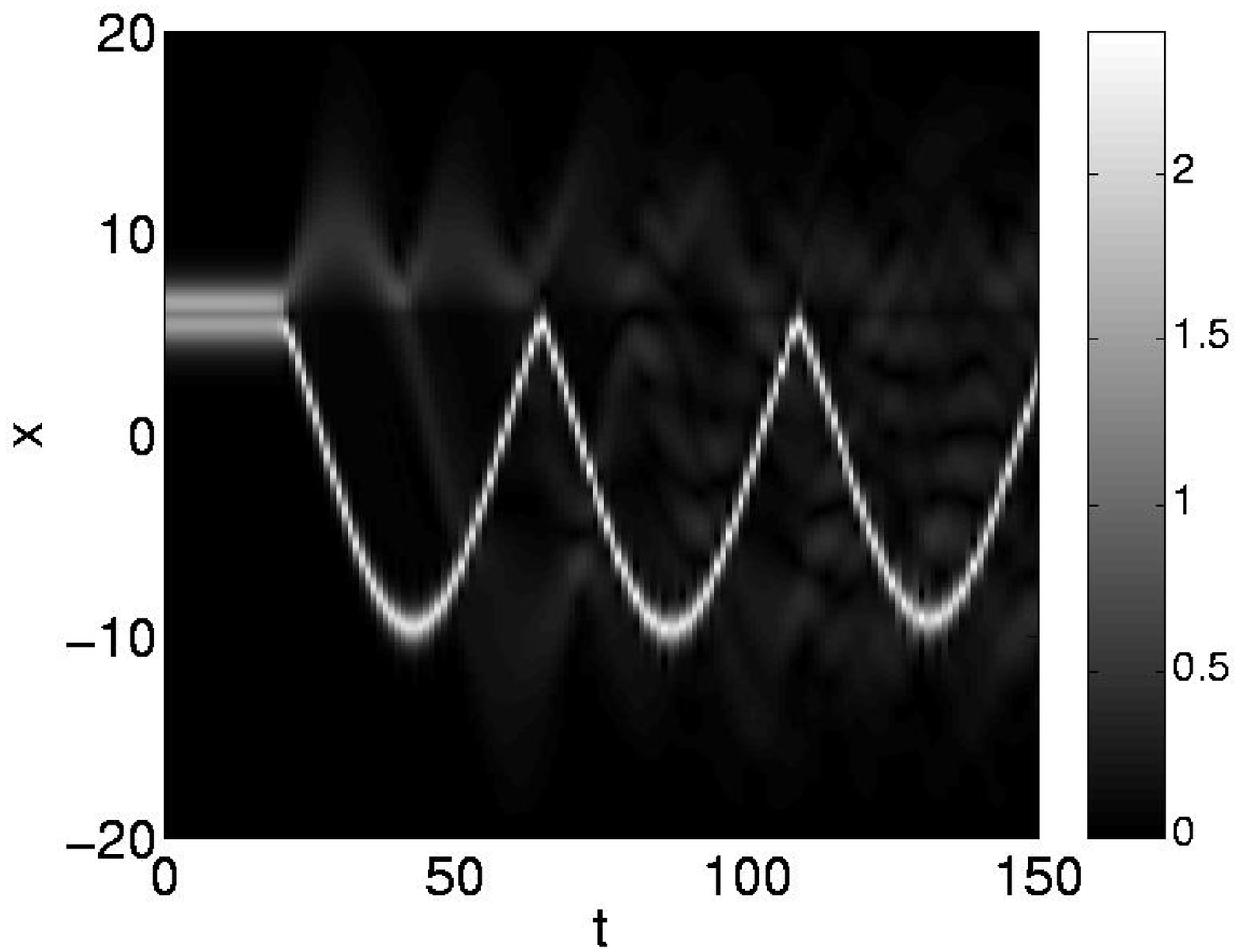} \includegraphics[width=5.cm,height=4cm,angle=0,clip]{\rootfig 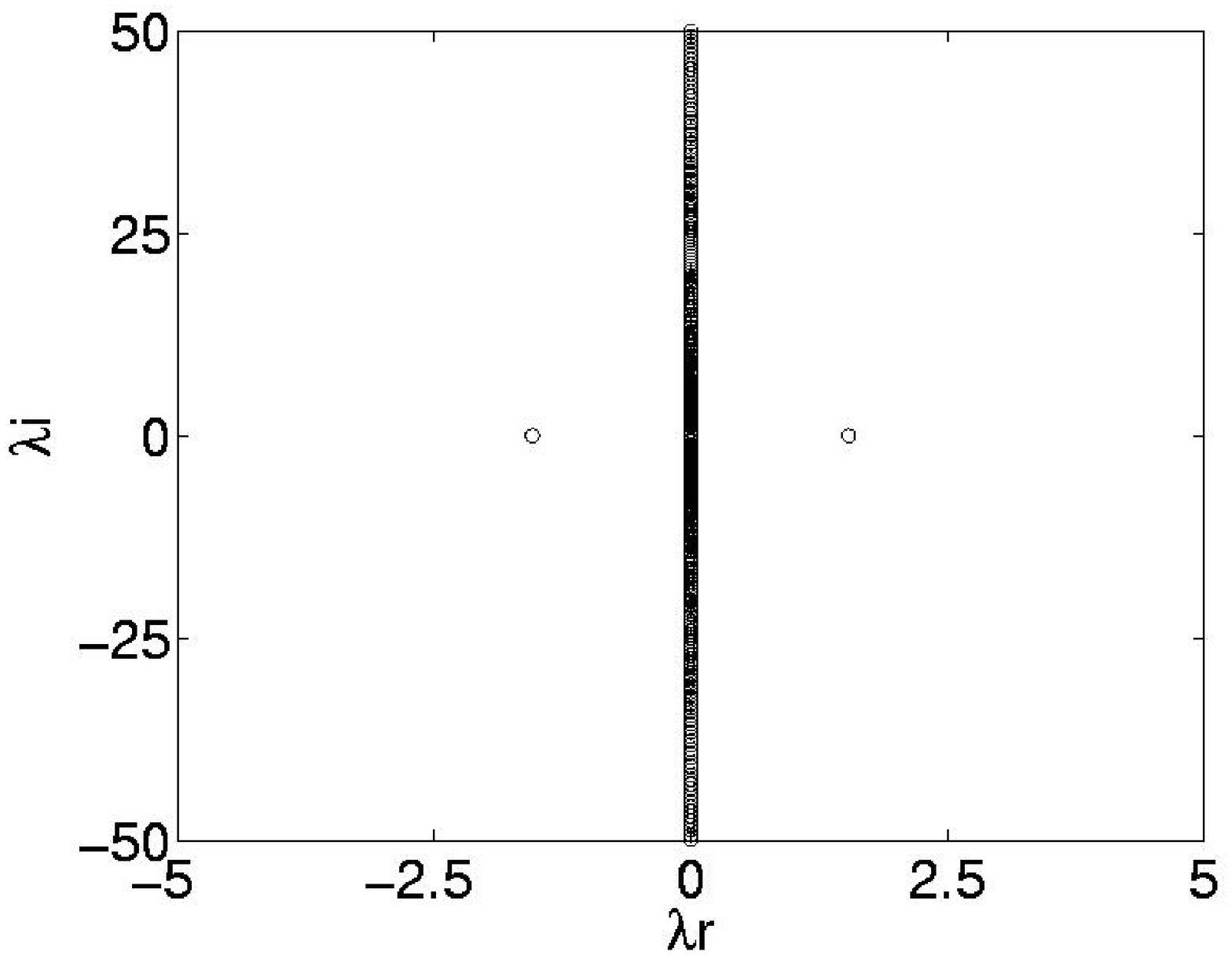} \\[0pt]
\includegraphics[width=5.cm,height=4cm,angle=0,clip]{\rootfig 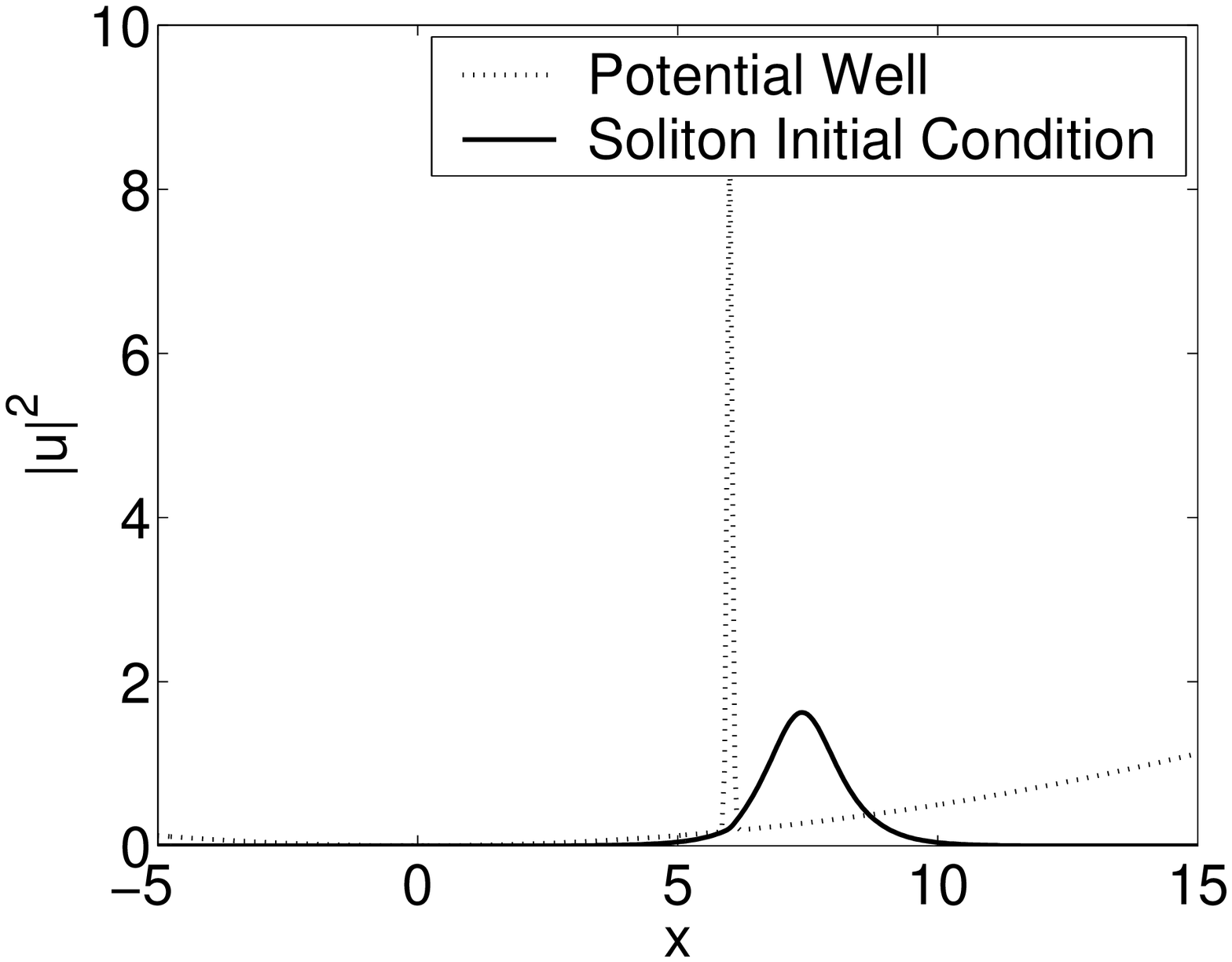}
\includegraphics[width=5.cm,height=4cm,angle=0,clip]{\rootfig
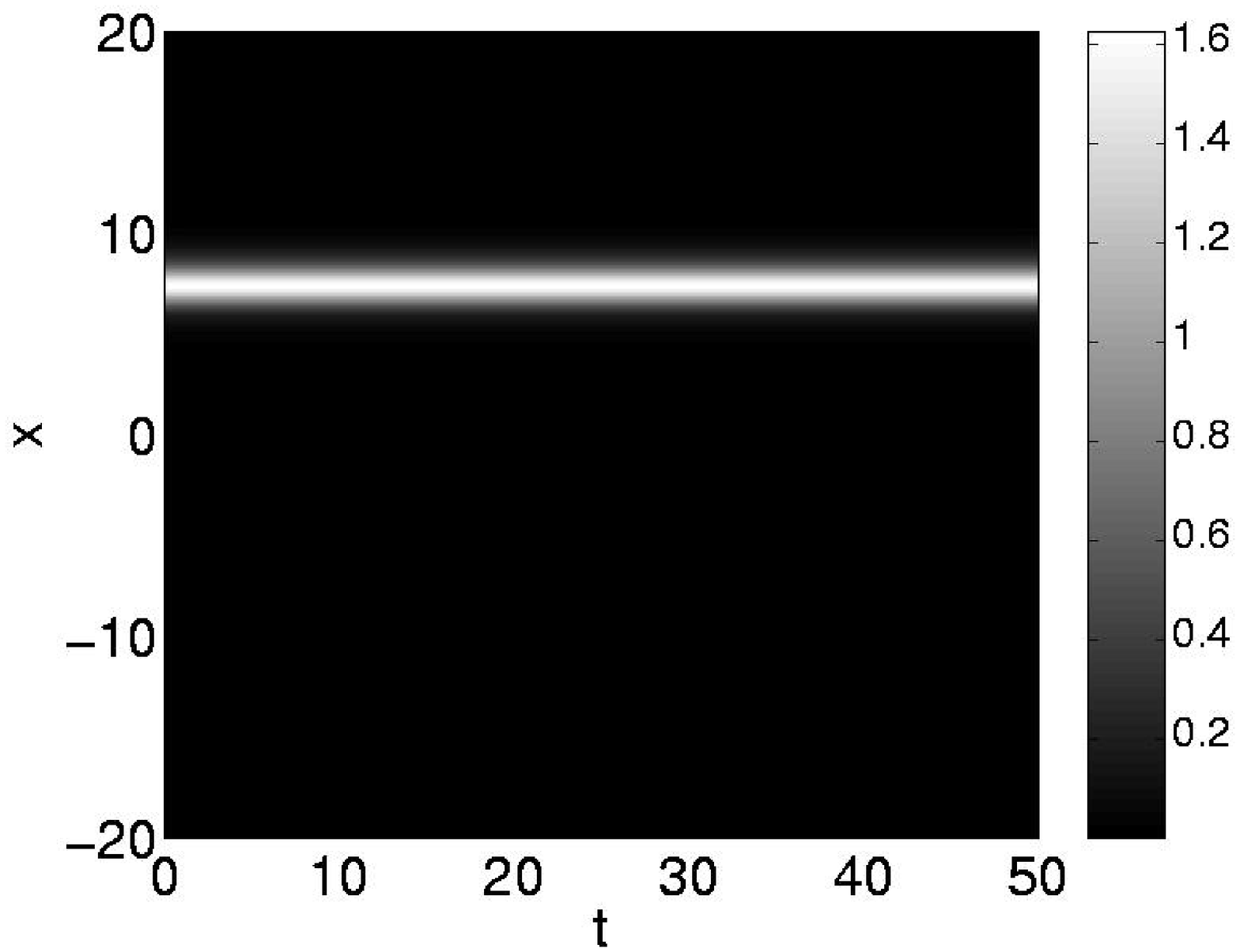}
\includegraphics[width=5.cm,height=4cm,angle=0,clip]{\rootfig
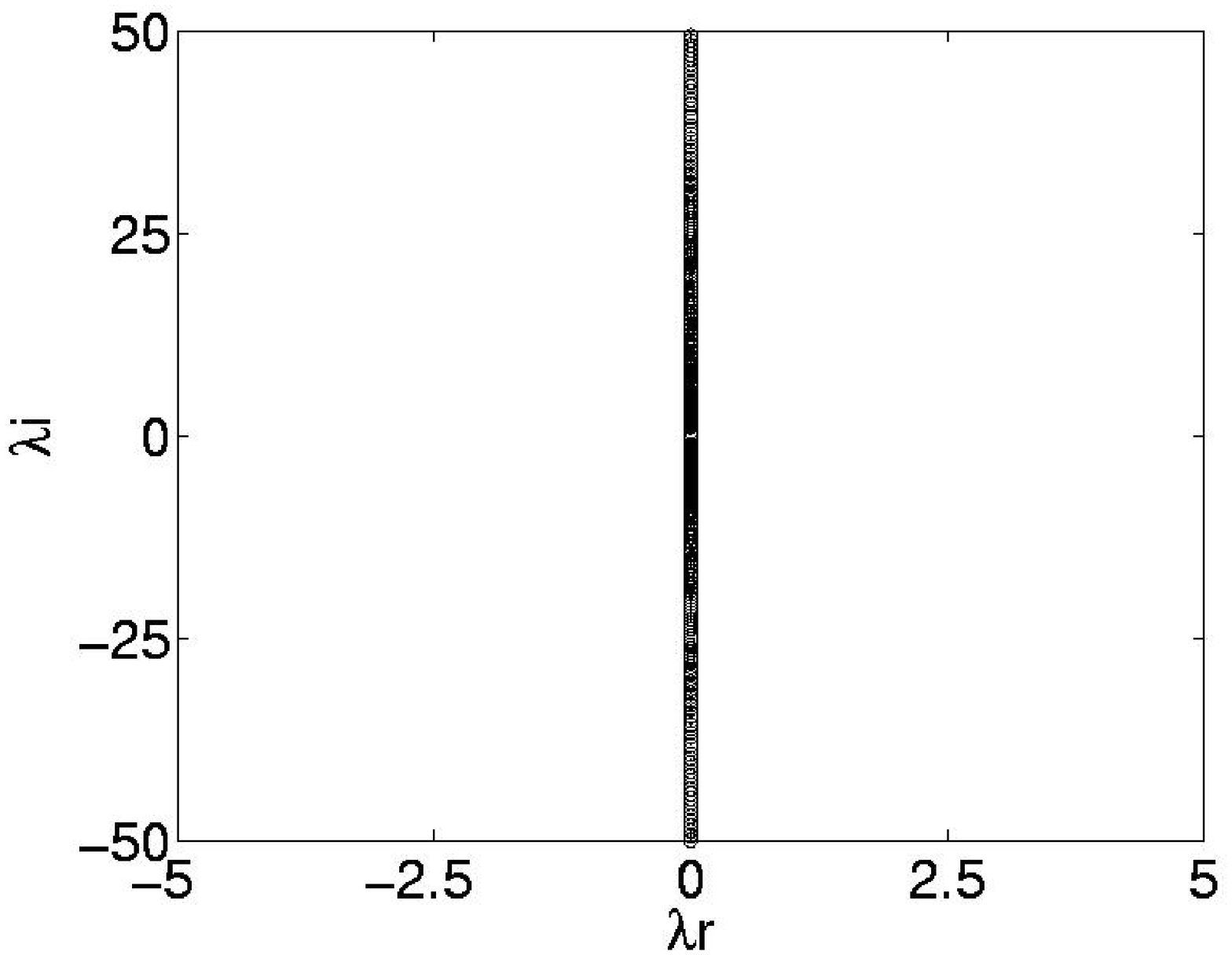}
\end{center}
\caption{Steady states of the bright soliton for the
repulsive defect: $V_{0}=-1$, $\protect\sigma =0.045$,
$\protect\eta =\protect\sqrt{2}$, $\Omega =0.1$ and $\protect\xi
=6$. The first row corresponds to the steady state at
$\protect\zeta =0$, the second row to the steady state centered at
the defect, and the third row to the steady state trapped to the
right of the impurity. For each row, the left panel displays the
numerically exact steady-state soliton profile, the middle panel
is the space-time evolution shown by means of contour plots, and
the right graph shows the spectral plane ($\protect\lambda
_{r},\protect\lambda _{i}$) for the stability eigenvalues
$\protect\lambda =\protect\lambda _{r}+i\protect\lambda _{i}$
corresponding to this solution. For the two stable steady states
(trapped at the defect and to the right of it), the solution
remains stationary as expected. On the other hand, for the
unstable steady state, after approximately 20 time units, the
instability fragments the soliton into a more localized part
oscillating to its right and a more extended part oscillating to
its left. } \label{Fig10}
\end{figure}

\begin{figure}[t]
\begin{center}
\includegraphics[width=5.cm,height=4cm,angle=0,clip]{\rootfig 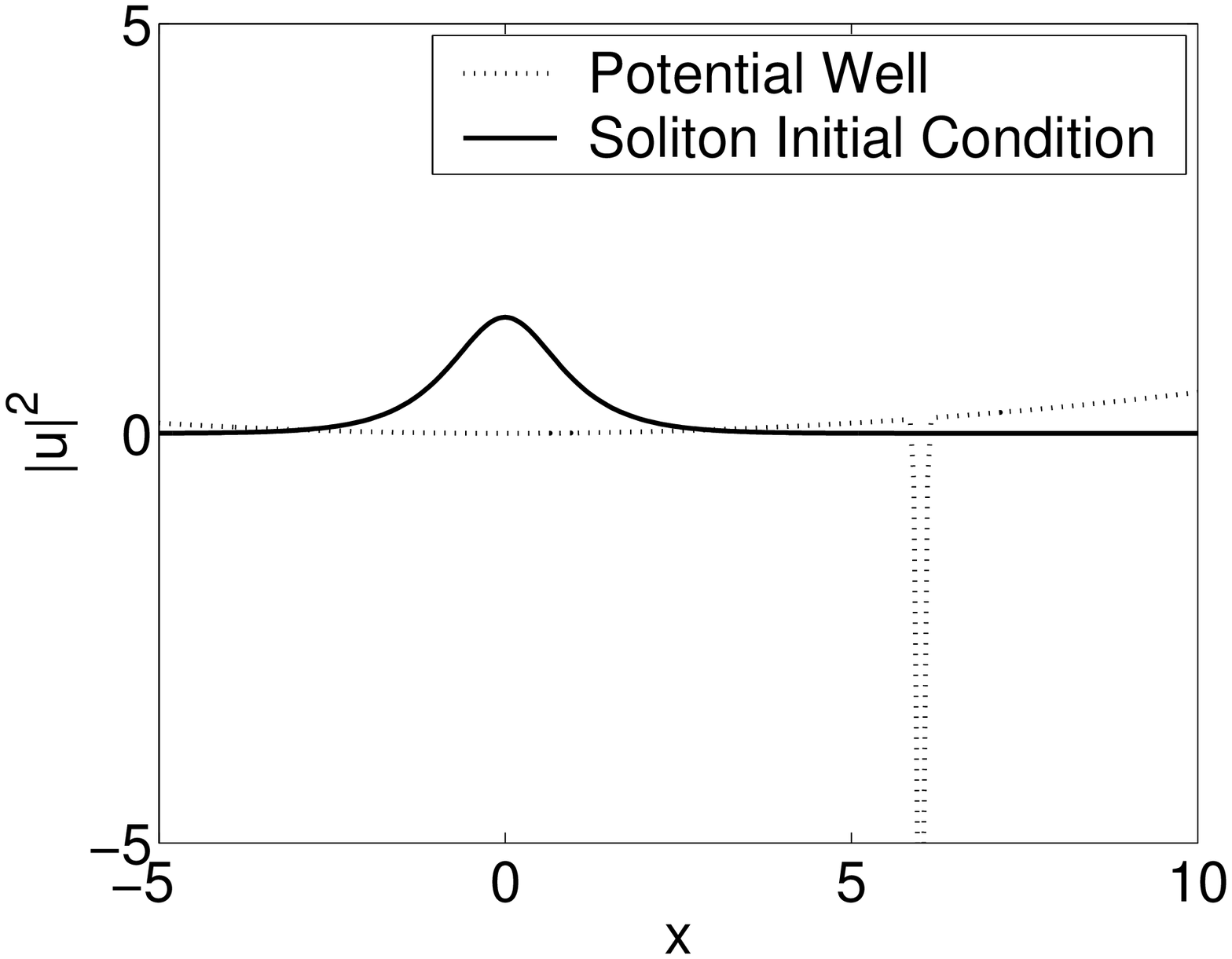}
\includegraphics[width=5.cm,height=4cm,angle=0,clip]{\rootfig
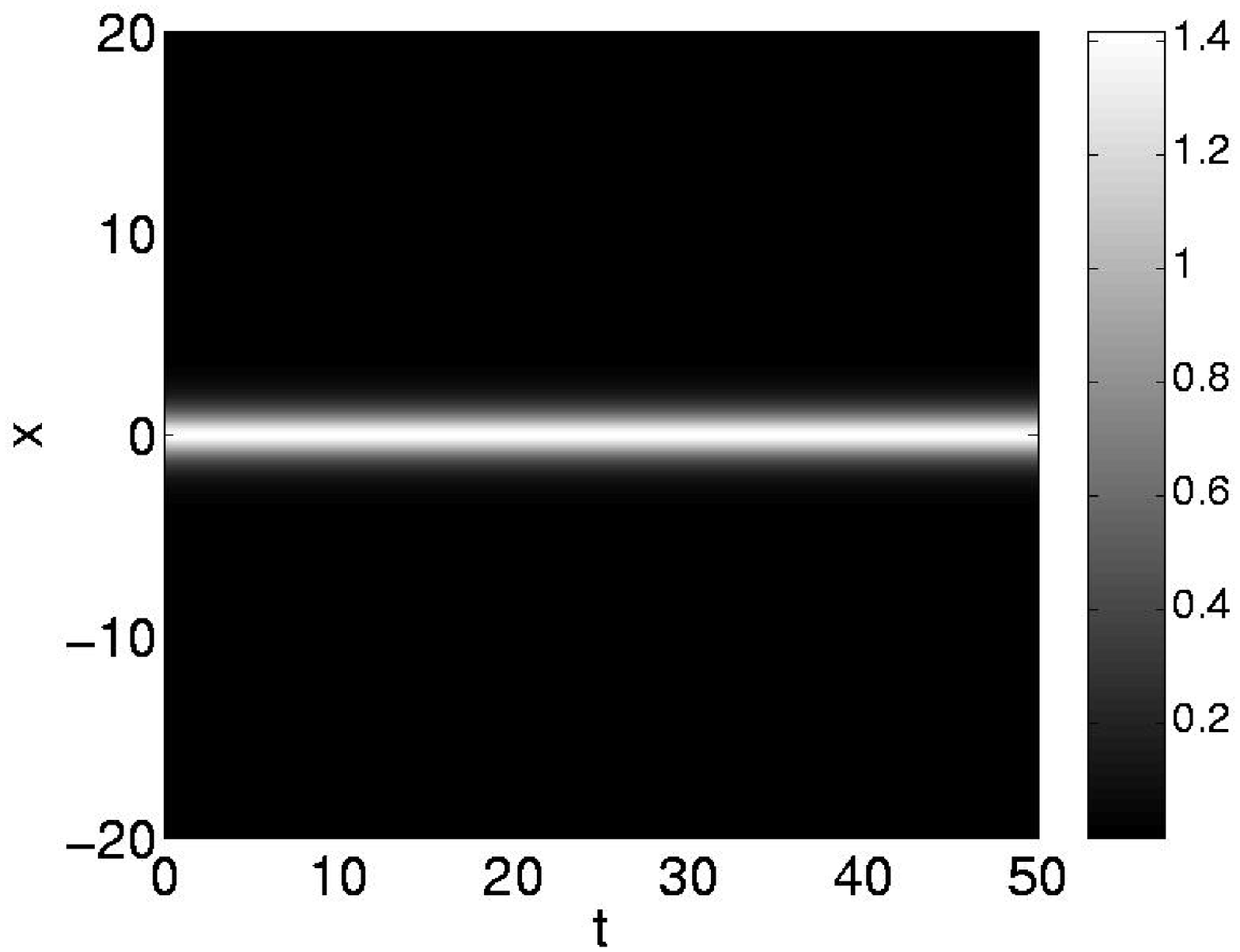} \includegraphics[width=5.cm,height=4cm,angle=0,clip]{\rootfig 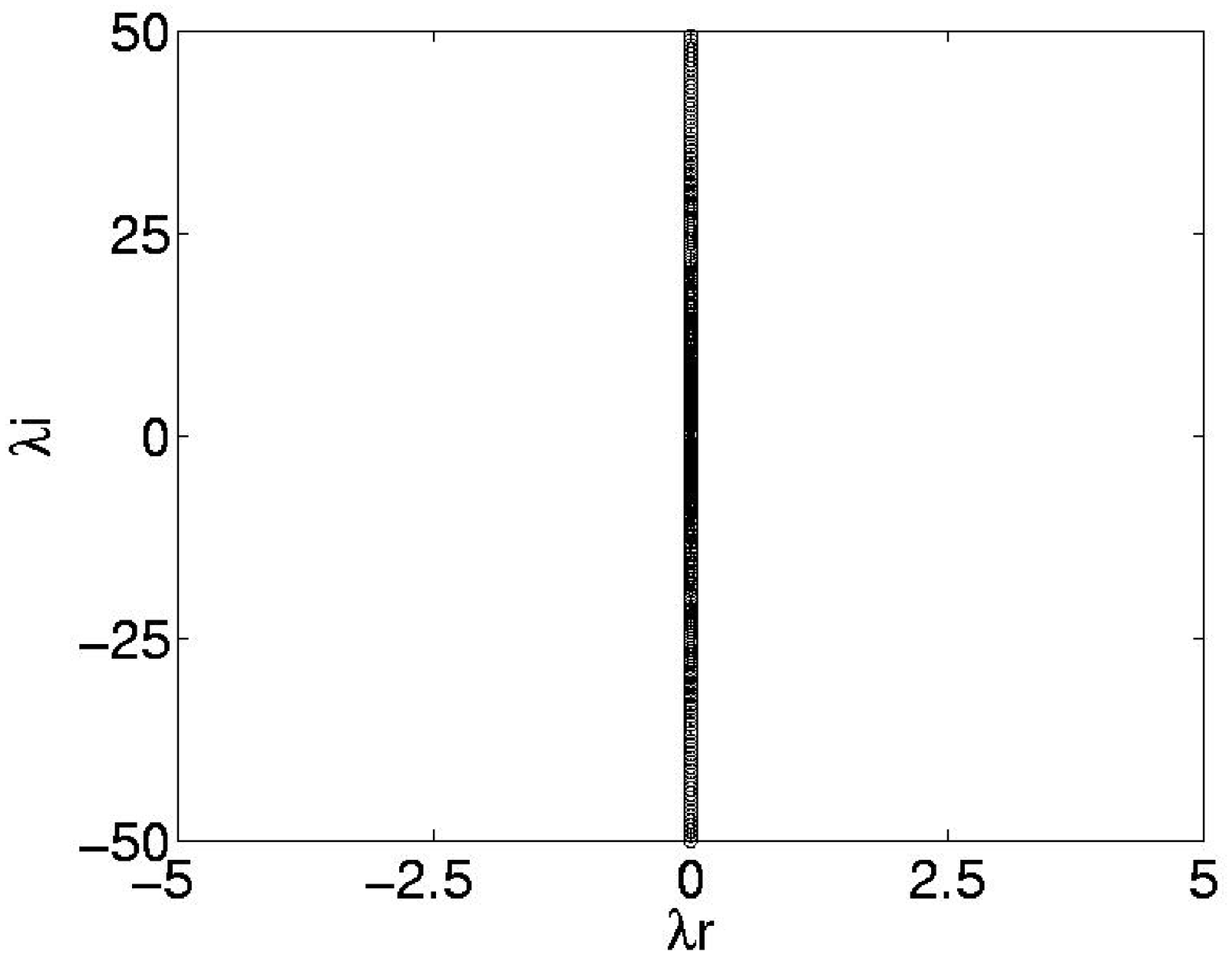} \\[0pt]
\includegraphics[width=5.cm,height=4cm,angle=0,clip]{\rootfig 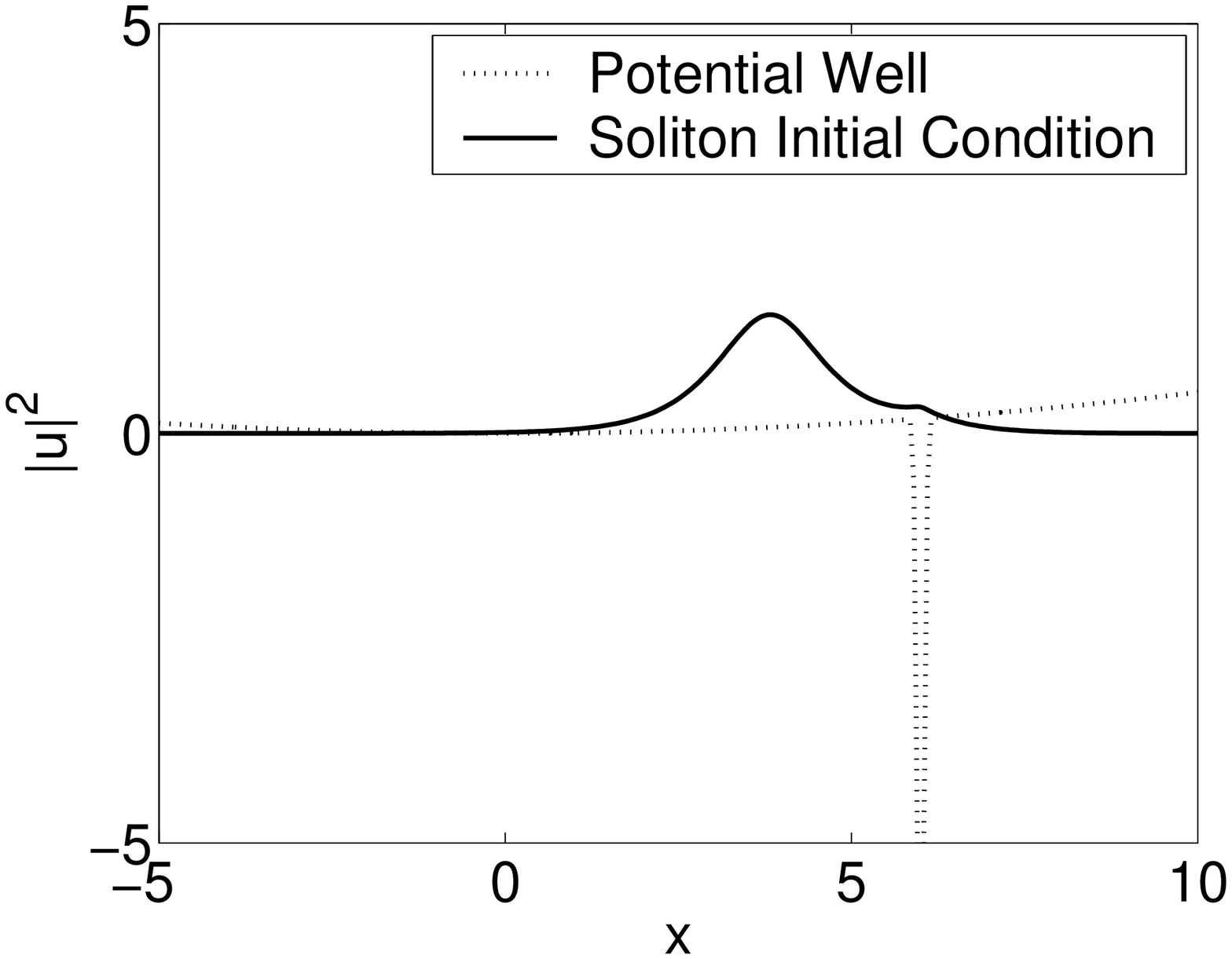}
\includegraphics[width=5.cm,height=4cm,angle=0,clip]{\rootfig
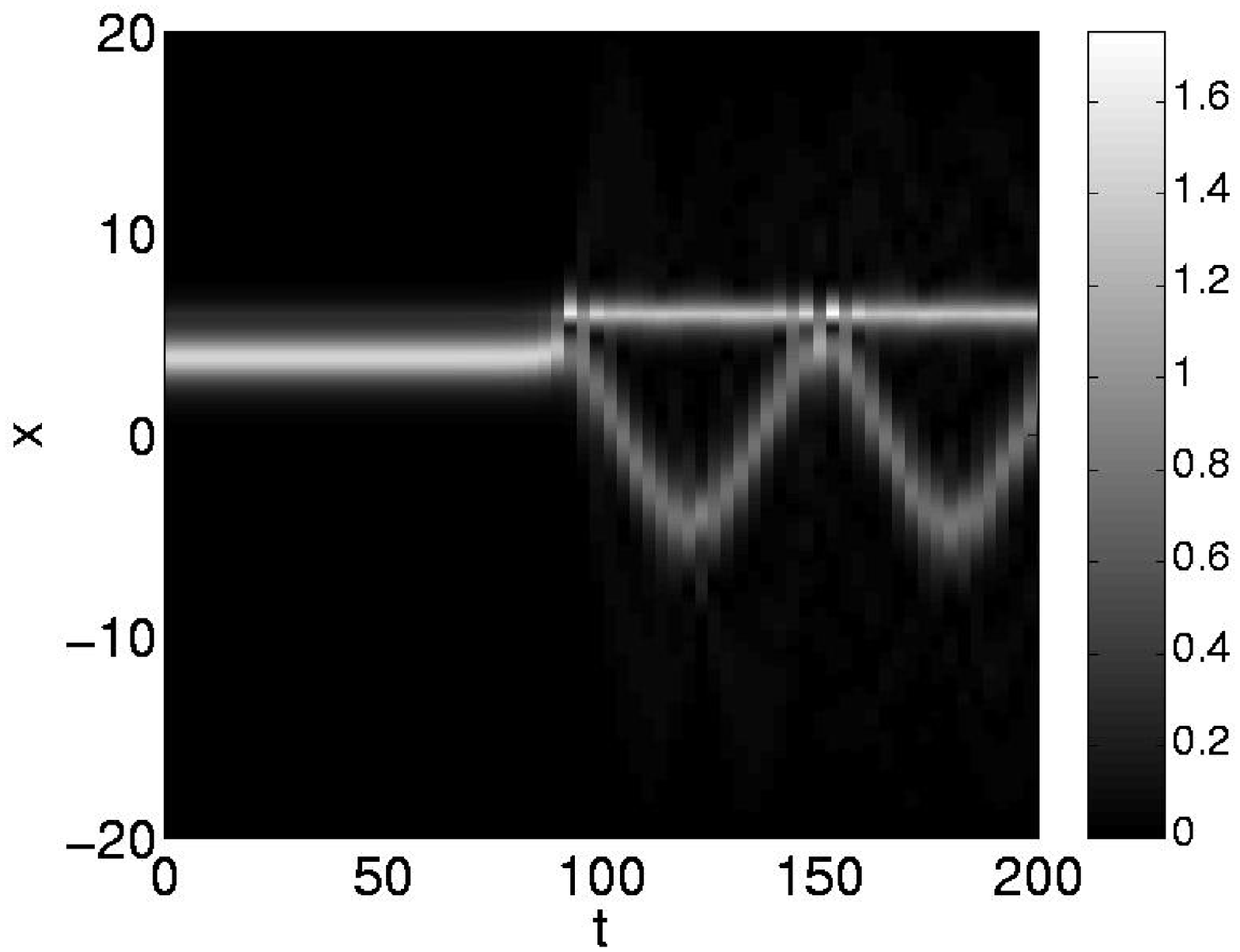} \includegraphics[width=5.cm,height=4cm,angle=0,clip]{\rootfig 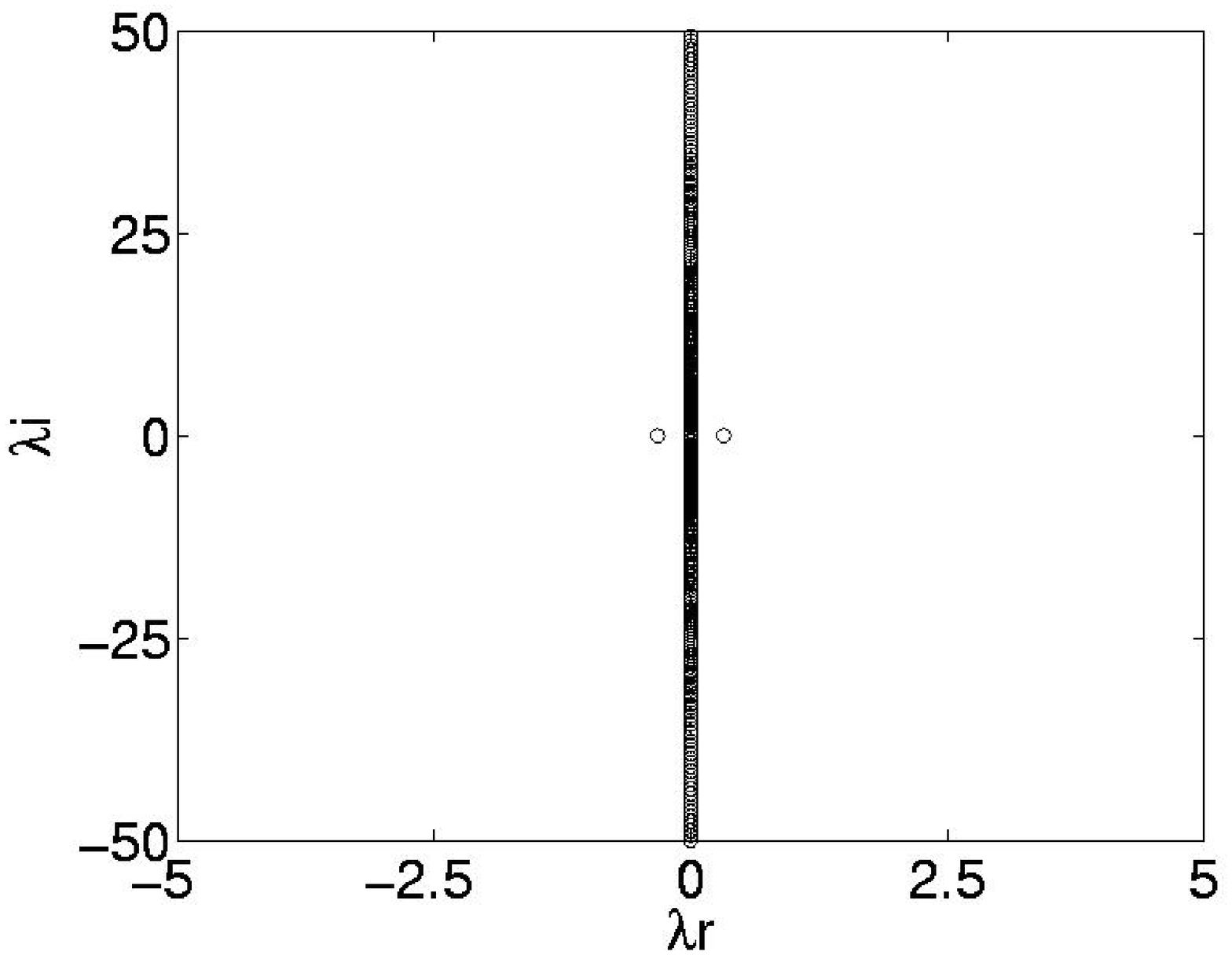} \\[0pt]
\includegraphics[width=5.cm,height=4cm,angle=0,clip]{\rootfig 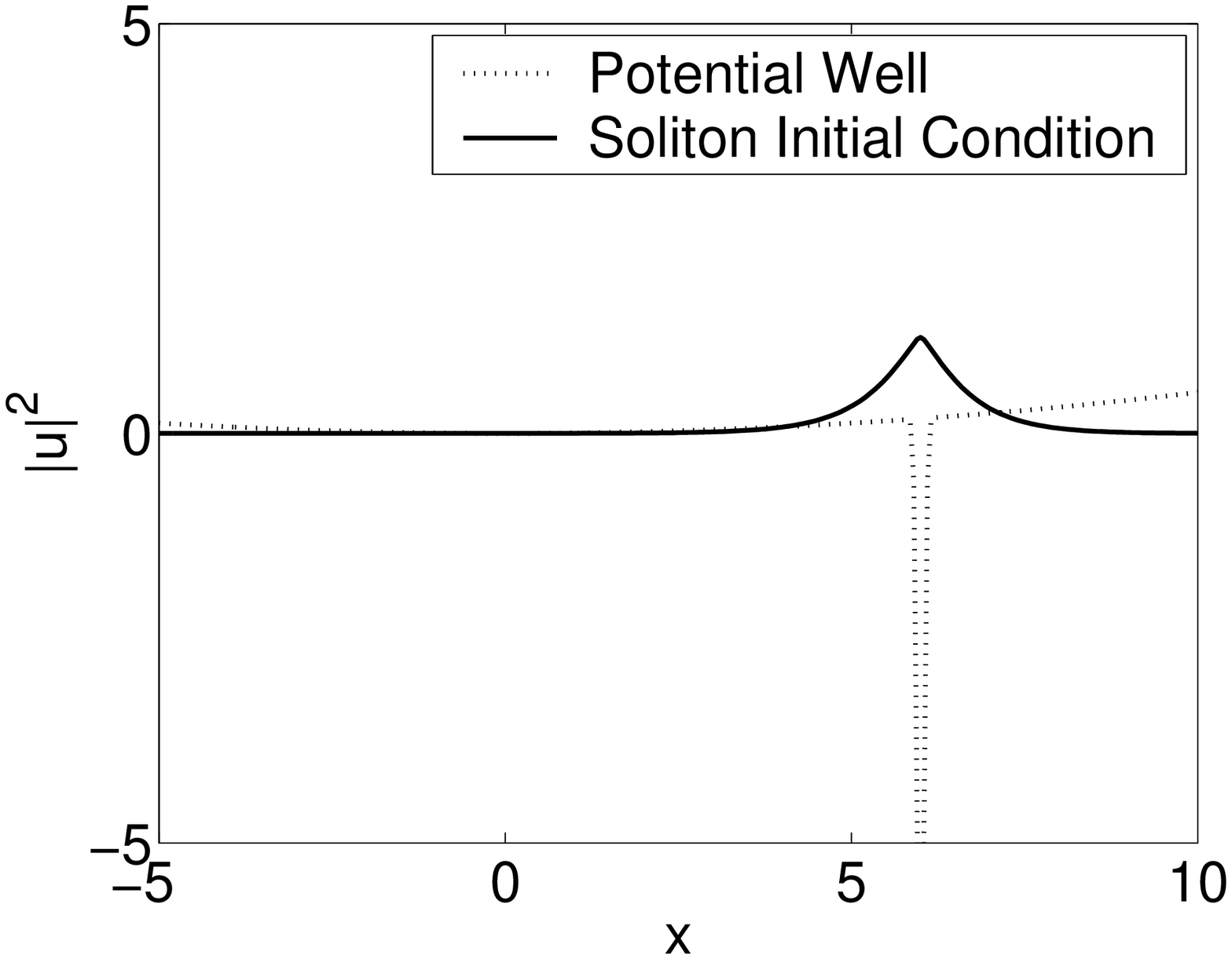}
\includegraphics[width=5.cm,height=4cm,angle=0,clip]{\rootfig
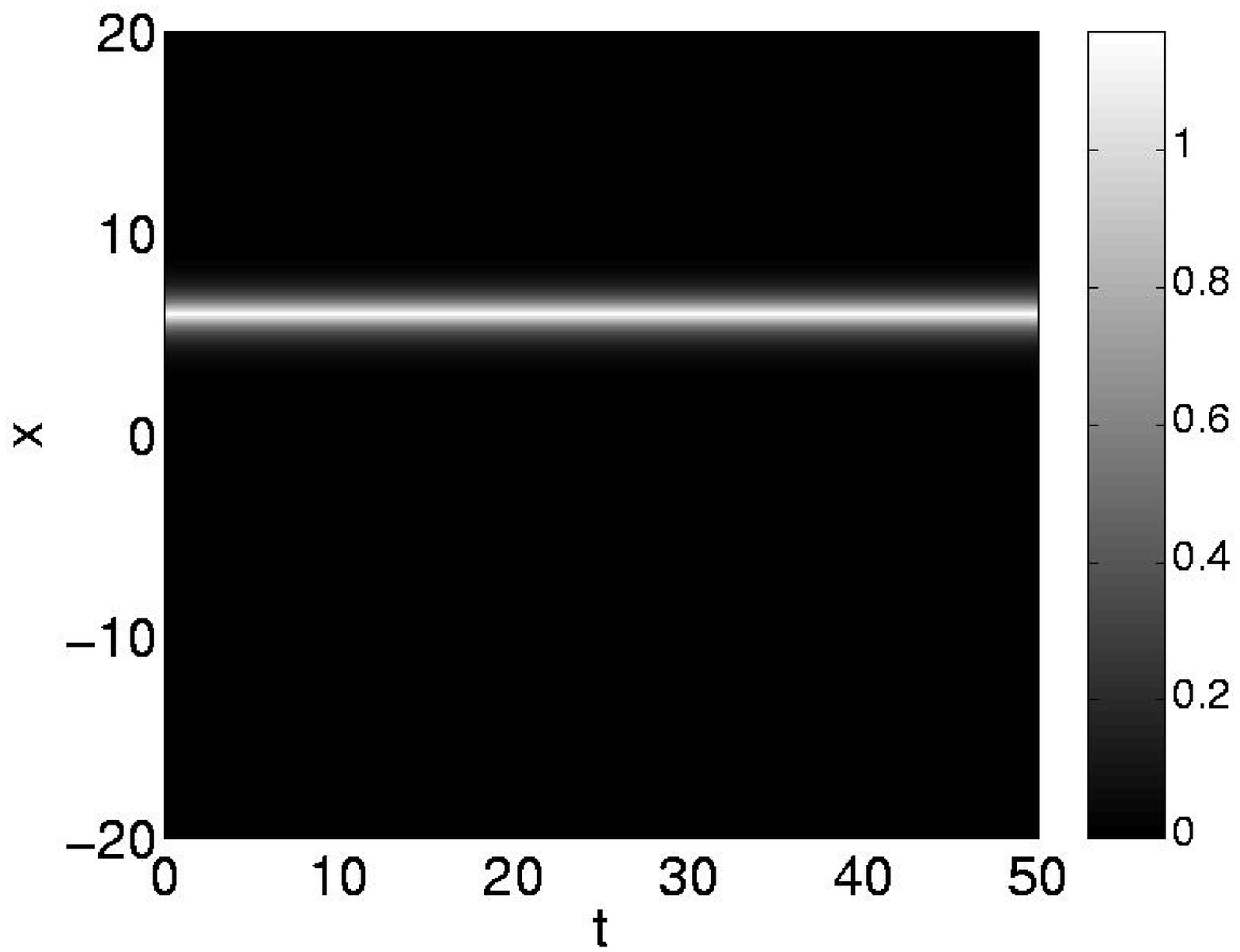}
\includegraphics[width=5.cm,height=4cm,angle=0,clip]{\rootfig
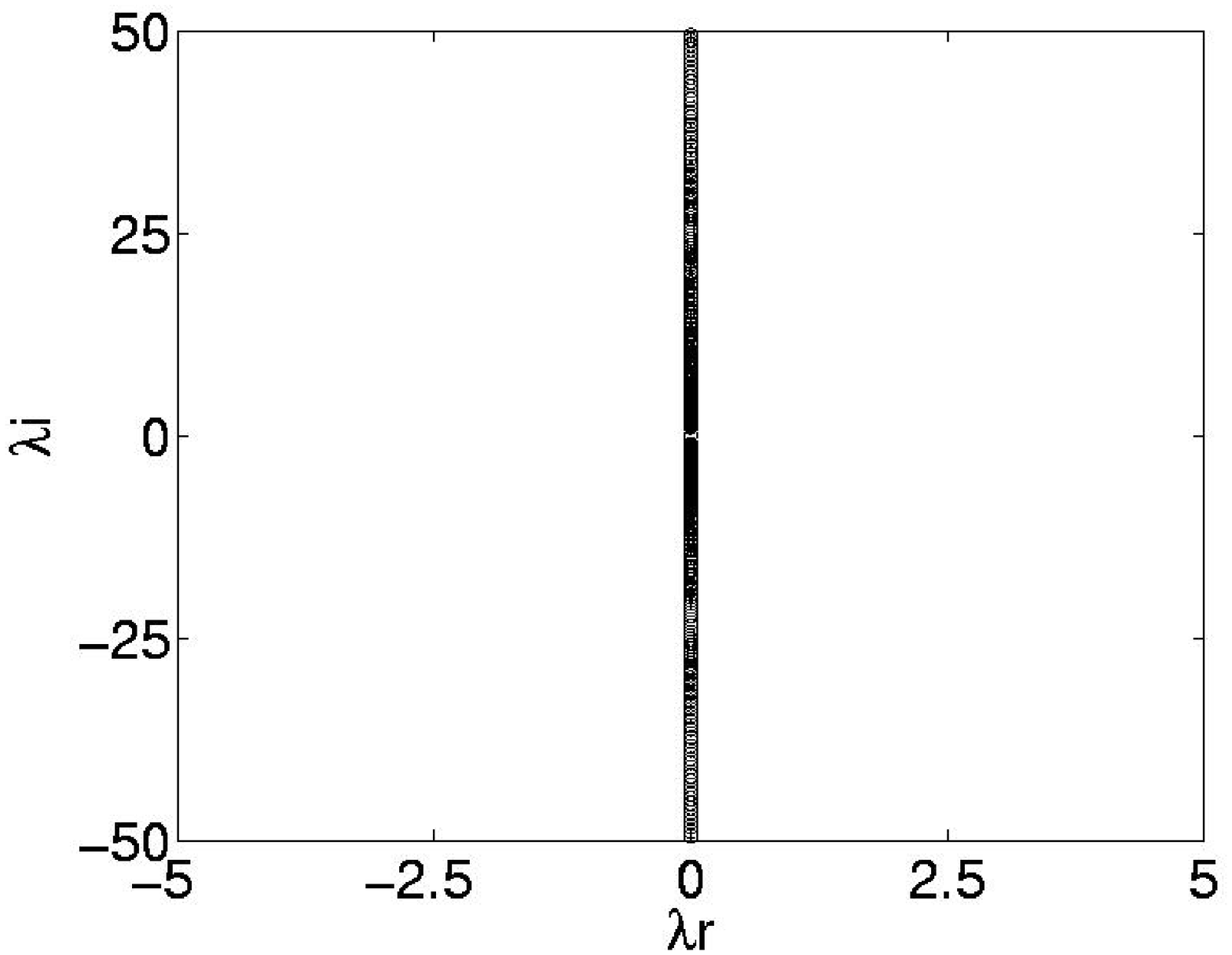}
\end{center}
\caption{Same as in the previous figure, but for an attractive
defect: $V_{0}=-1$, $\protect\sigma =0.045$, $\protect\eta
=\protect\sqrt{2}$, $\Omega =0.1$ and $\protect\xi =6$. Notice
that now the unstable steady state is to the left of the
attractive defect, while its unstable time-evolution leads to its
trapping at the defect.} \label{Fig11}
\end{figure}

The qualitative predictions about the nature of the steady states and their
stability have been tested for repulsive and attractive defects, as shown in
Figs.\ \ref{Fig10} and \ref{Fig11}, respectively. 
In these figures, three left panels show the spatial profiles of
the stable branch at $\zeta =0$, and the unstable and stable
branches in the neighborhood of the defect. The middle panels show
the temporal evolution of each one of these solutions, while the
right panels show the results of the linear stability analysis.
The latter set clearly illustrates the instability of the middle
branch due to the presence of a real eigenvalue pair. It is also
noteworthy that, in the case of the repulsive defect (in which
case an unstable solution is centered at the defect) the soliton
oscillates around the nearby stable steady state, shedding
radiation waves, cf.\ Fig.\ \ref{Fig10}. On the other hand, in the
attractive case the unstable solution centered beside the defect
is ``captured" by the defect, resulting in its trapping at the
defect's center, cf.\ Fig.\ \ref{Fig11}. However, a fraction of
the condensate is also emitted from the defect in the process,
leading to oscillations that can be observed in the respective
space-time-evolution panel.

\begin{figure}[t]
\begin{center}
\includegraphics[width=5.6cm,height=5cm,angle=0,clip]{\rootfig 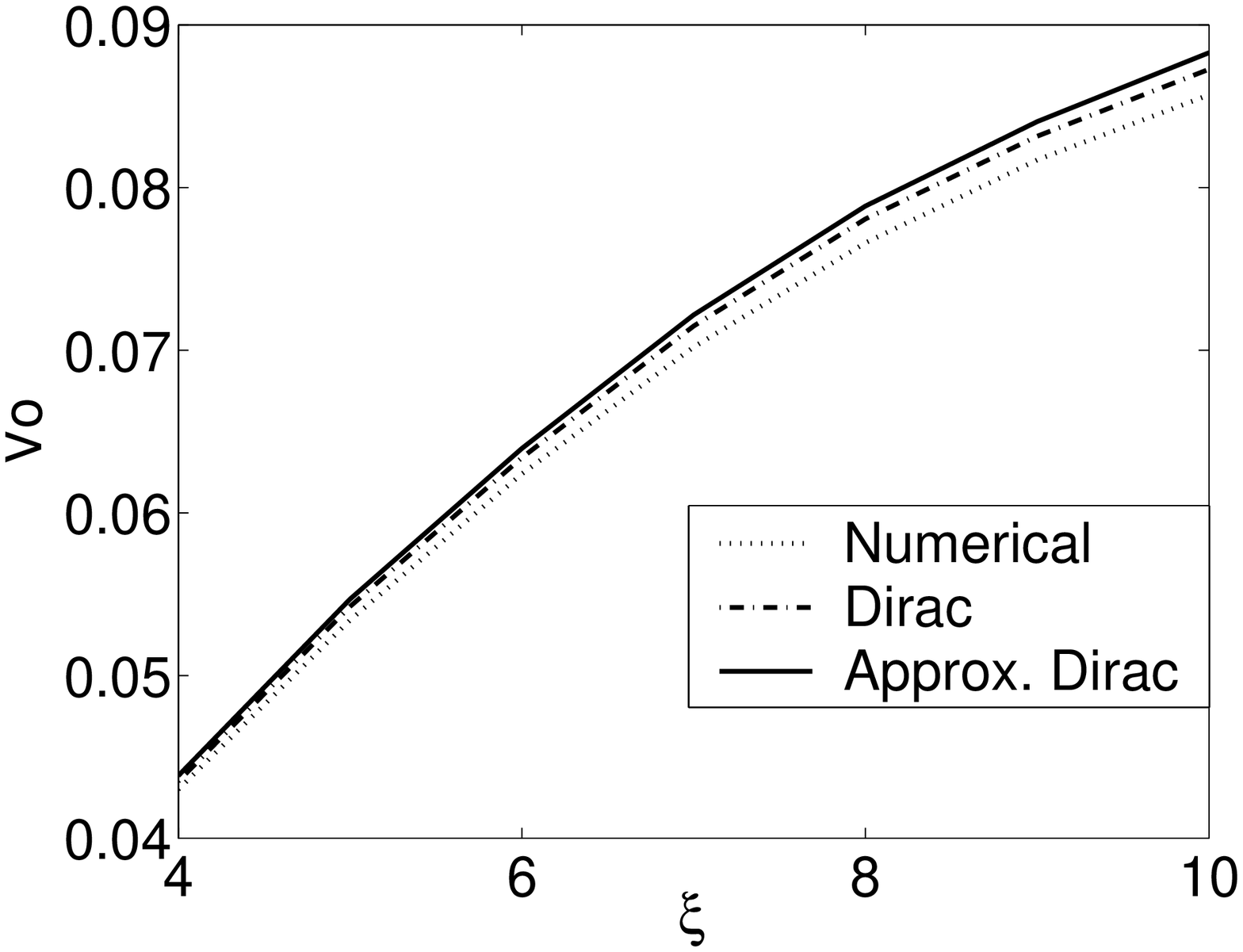}
\includegraphics[width=5.6cm,height=5cm,angle=0,clip]{\rootfig 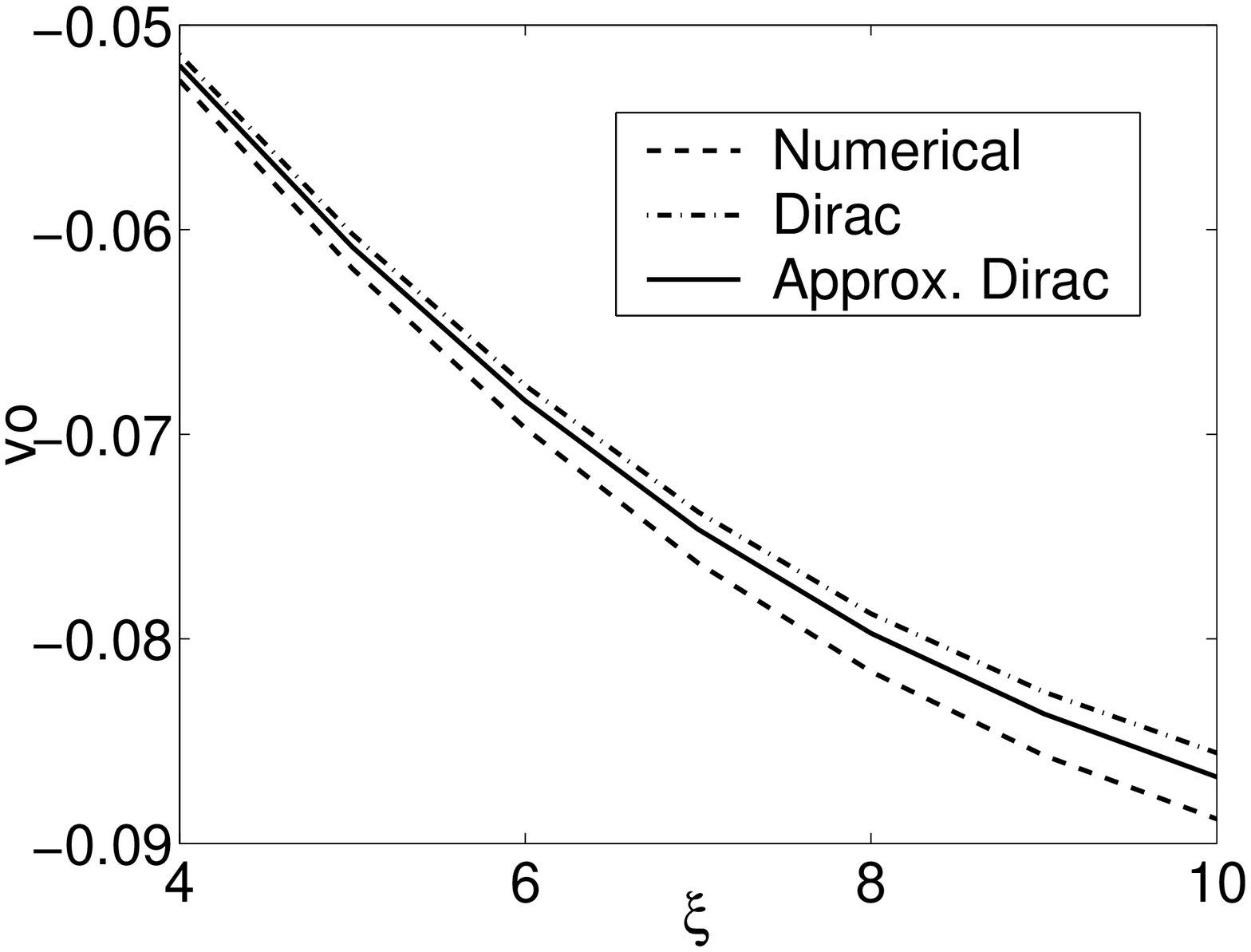}
\end{center}
\caption{Critical values of $V_{0}$, corresponding to the
disappearance of two steady states, for $\protect\sigma =0.045$.
The left and right graphs pertain to the attractive and repulsive
defect, respectively.} \label{FigBif}
\end{figure}

To verify the analytical results following from Eqs. (\ref{meq8})
and (\ref{meq9}), we have compared the analytically predicted
critical value of $V_{0} $ (for which a double root appears) with
the numerically obtained turning point for the saddle-node
bifurcation. This comparison was performed for many values of the
impurity center $\xi $. In fact, the critical value was predicted
using two different forms of the analytical prediction: one with
the Dirac $\delta $-function proper, and another one with the Gaussian
approximation for the $\delta $-function and an accordingly
modified version of Eq.\ (\ref{meq9}), namely\begin{equation}
F_{\mathrm{imp}}+F_{\mathrm{trap}}=\int\limits_{-\infty }^{\infty
}V(x)\frac{\partial }{\partial x}\left( \left\vert u(x)\right\vert
^{2}\right) dx=0, \label{ModEqn}
\end{equation}
with the function $V(x)$ incorporating the parabolic magnetic trap
and the Gaussian impurity terms. 
Here the
integration was performed with the numerically implemented $V(x)$,
and the best fit of $u(x)$ to a hyperbolic secant waveform has
been used in Eq.\ (\ref{ModEqn}). The parameter values along with
the resulting critical values of $V_{0}$ are given in Fig.\
\ref{FigBif}. In all cases, the numerical results for the
bifurcation point closely match the theoretical predictions.

\begin{figure}[t]
\begin{center}
\includegraphics[width=5.0cm,height=4cm,angle=0,clip]{\rootfig
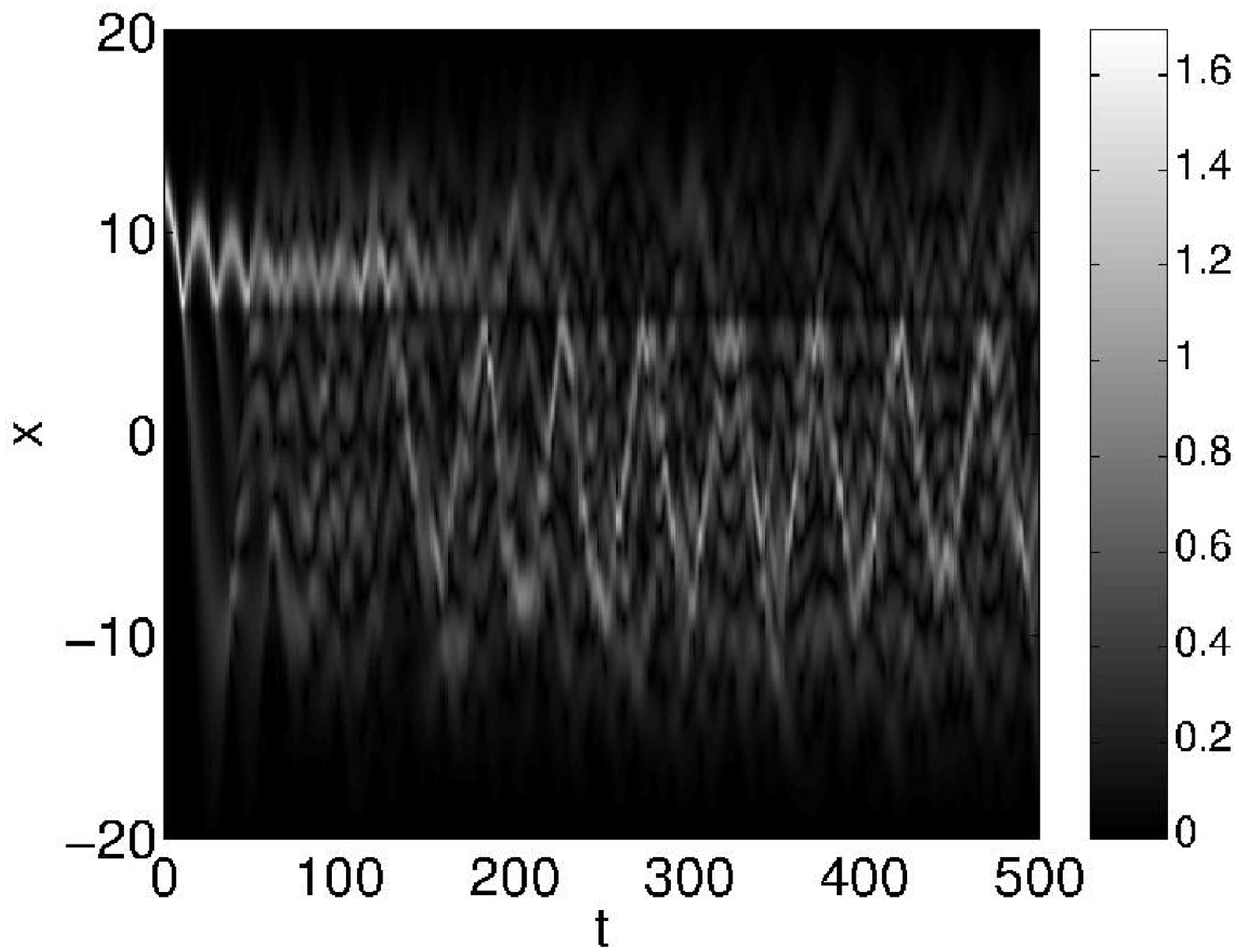}
\includegraphics[width=5.0cm,height=4cm,angle=0,clip]{\rootfig
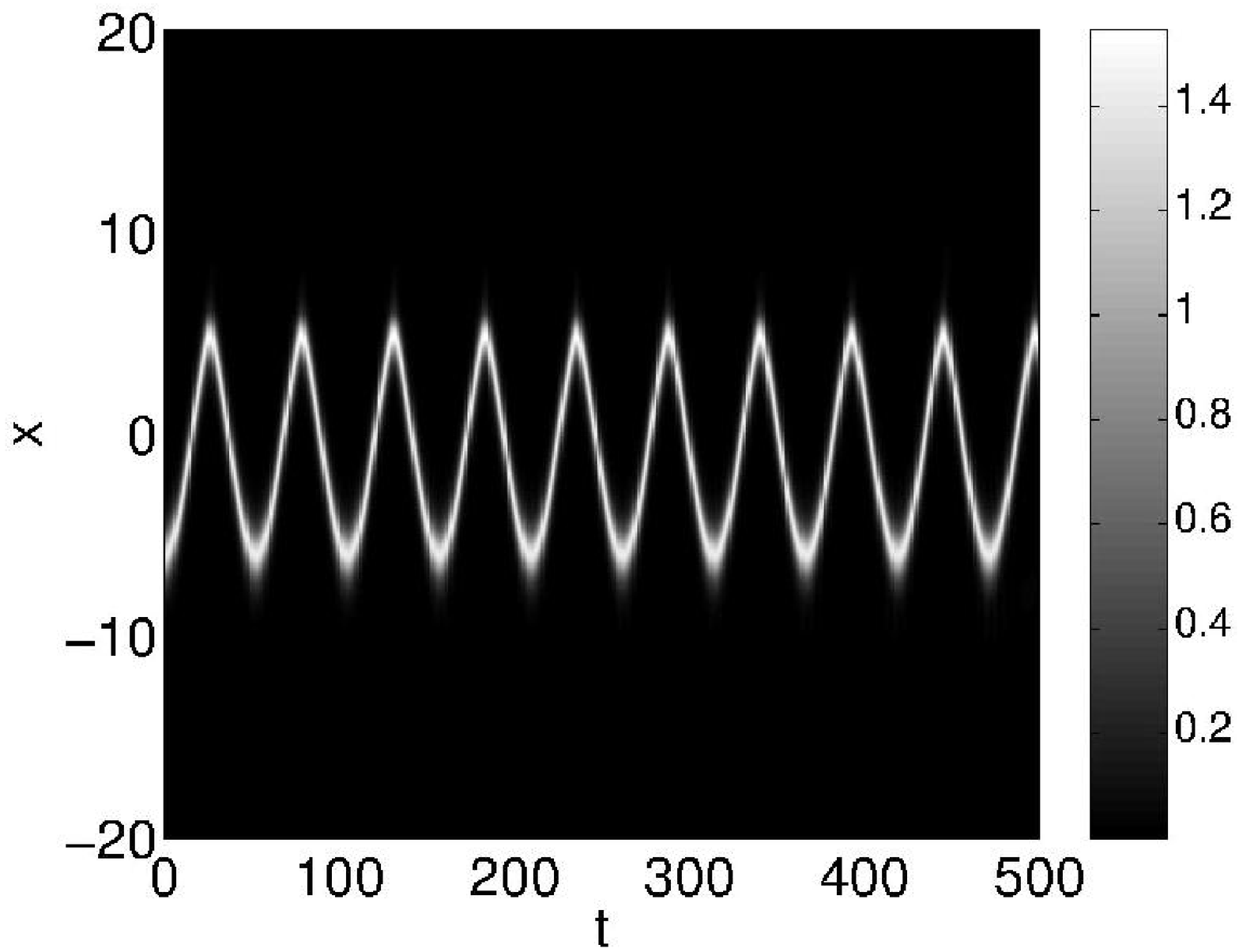}
\includegraphics[width=5.0cm,height=4cm,angle=0,clip]{\rootfig
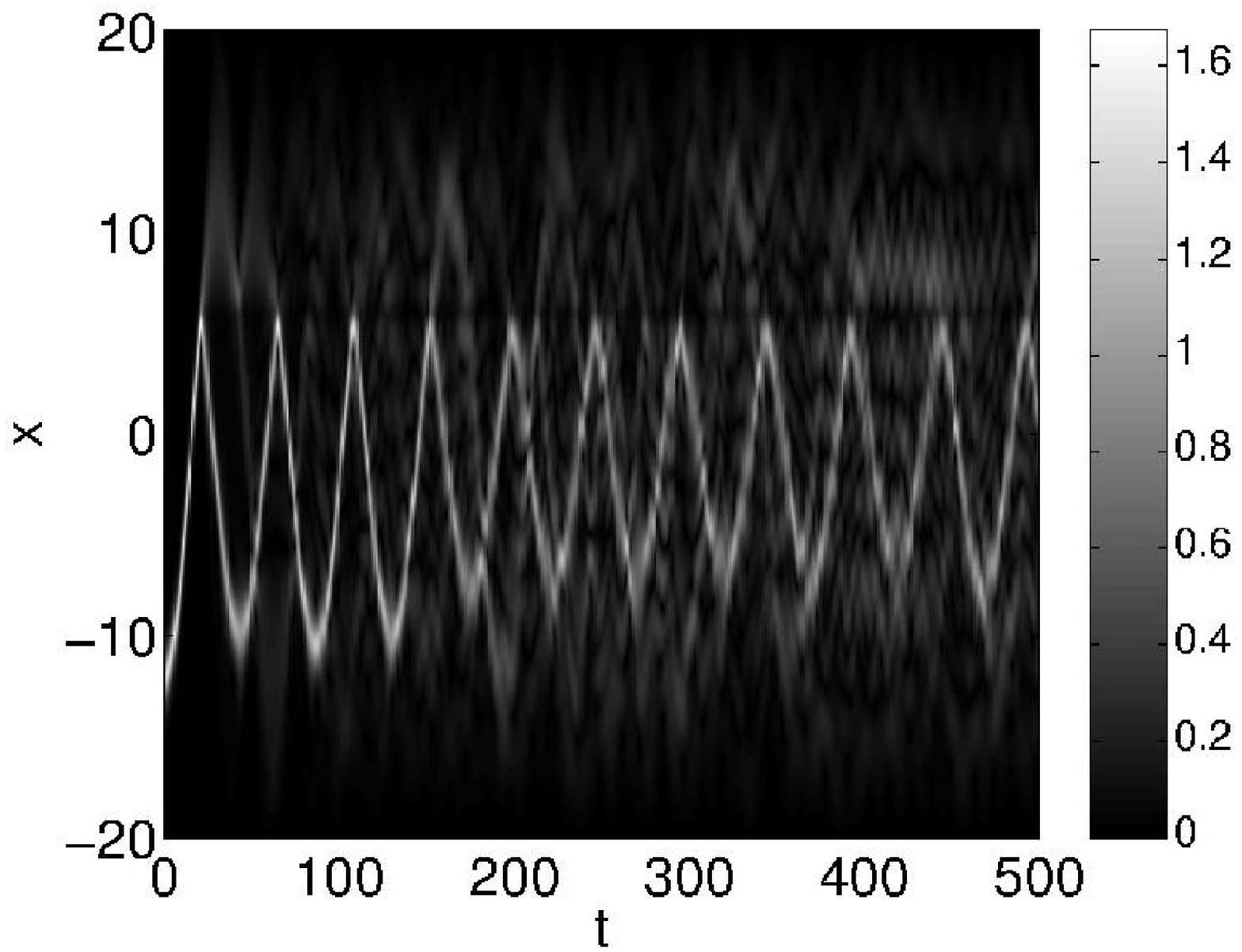} \\[0pt]
\includegraphics[width=5.0cm,height=4cm,angle=0,clip]{\rootfig
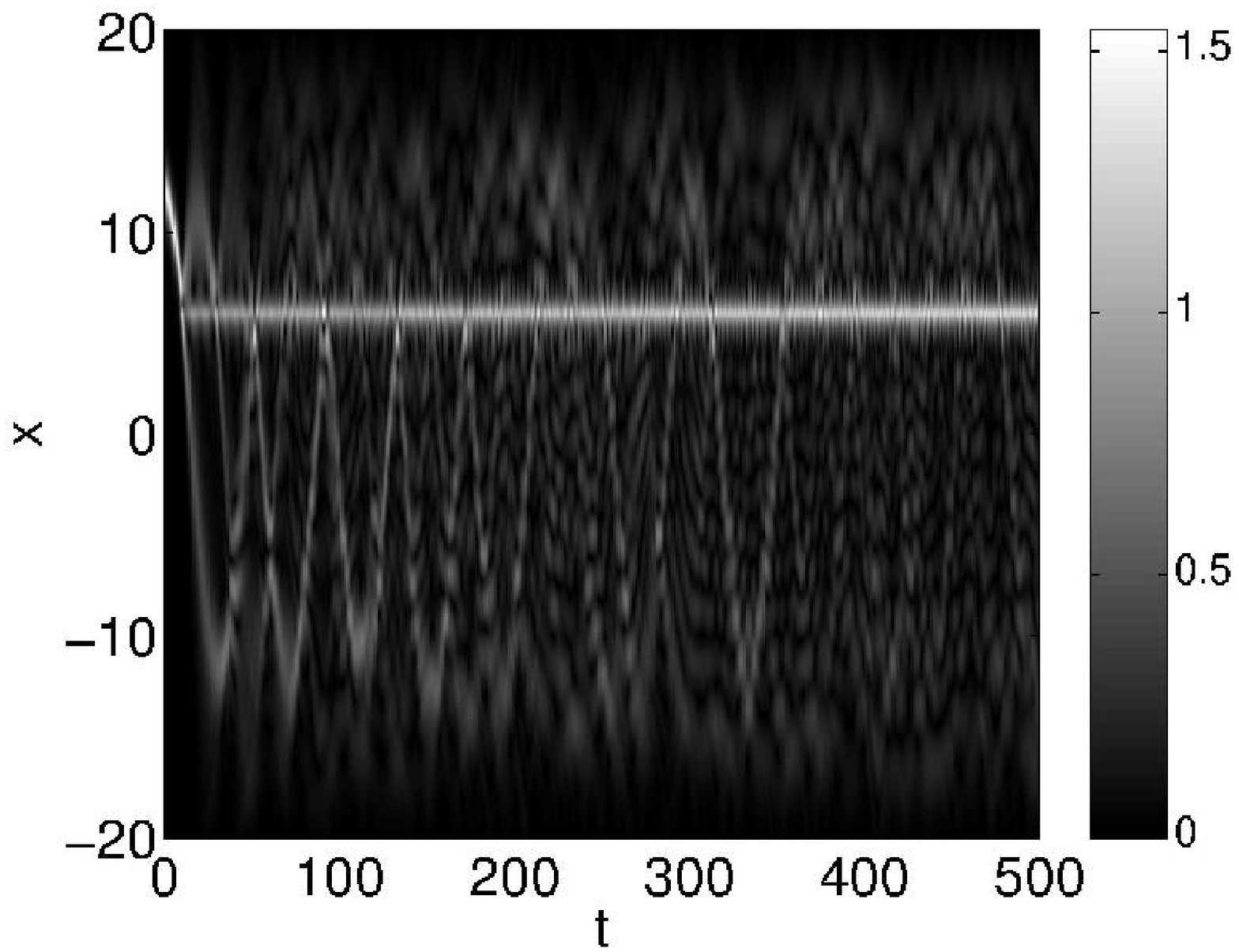}
\includegraphics[width=5.0cm,height=4cm,angle=0,clip]{\rootfig
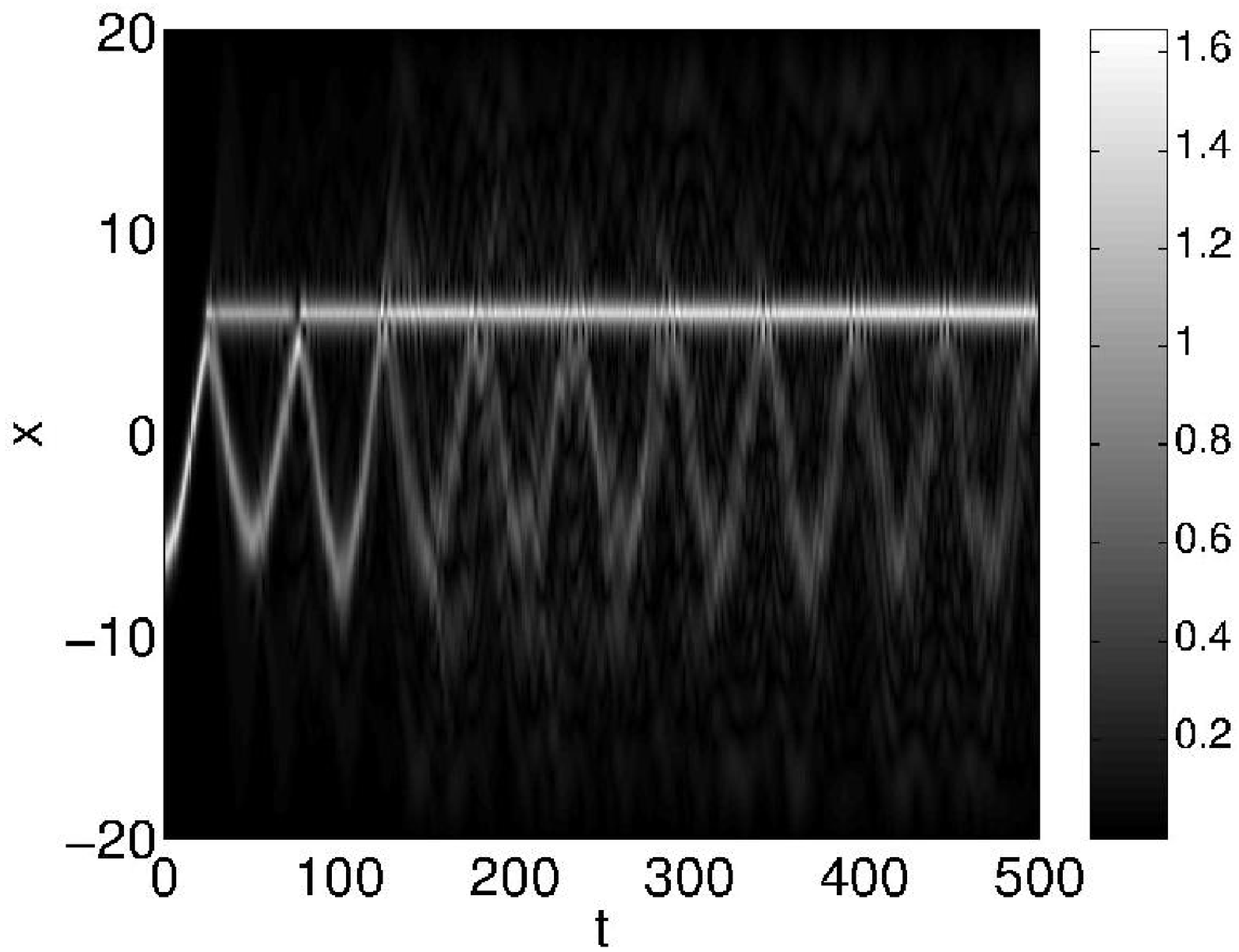}
\includegraphics[width=5.0cm,height=4cm,angle=0,clip]{\rootfig
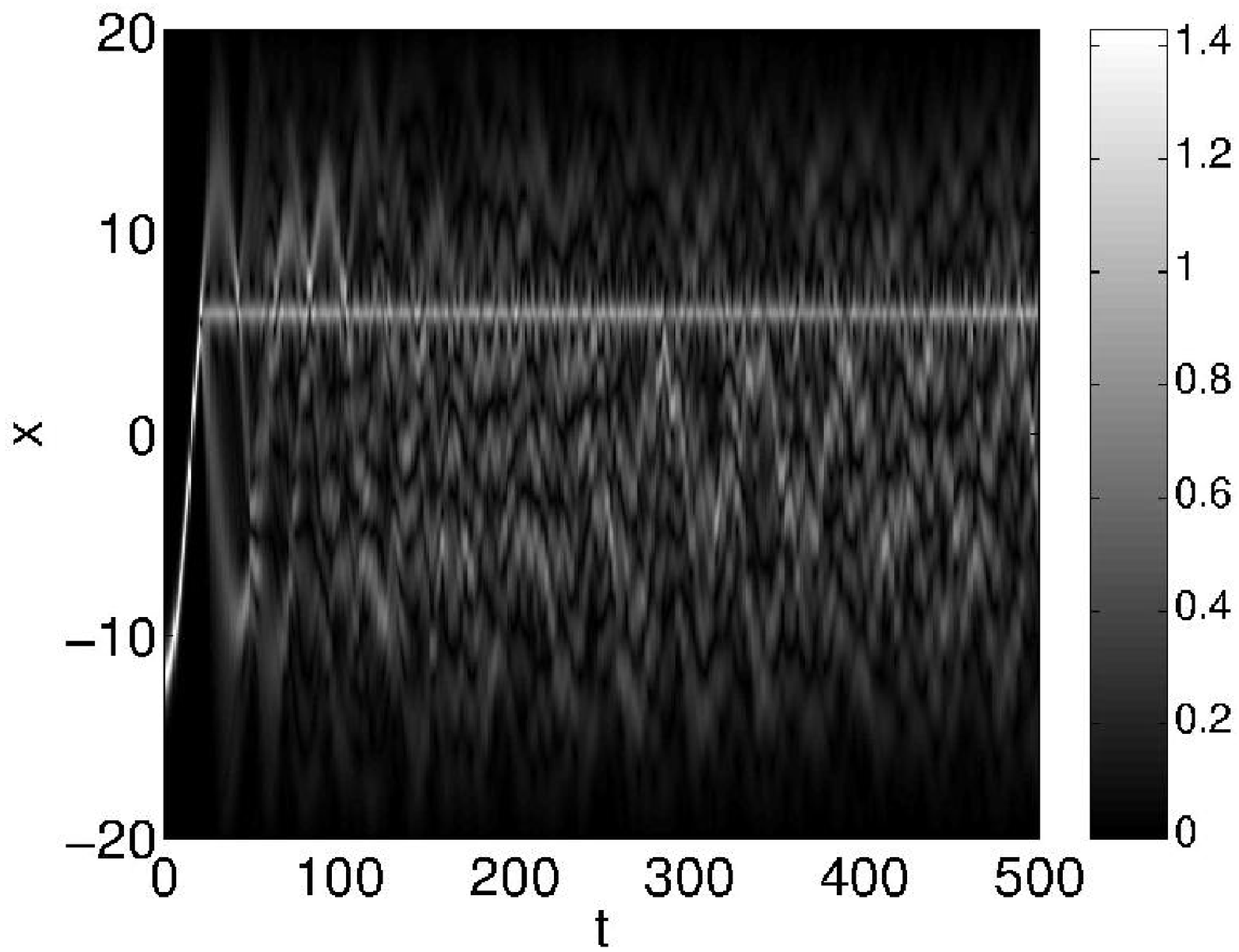}
\end{center}
\caption{Examples of the soliton interaction with the repulsive
defect (top panels) and the attractive one (bottom panels) located
at $\protect\xi =6$. The soliton is initially offset with respect
to the steady states of the model. For all cases, the parameters
are $V_{0}=\pm 1$, $\protect\sigma =0.045$ and $\protect\eta
=\protect\sqrt{2}$. The initial position of the soliton is
$\protect\zeta _{0}=12,-6,-12$, for the plots from left to right.
In the repulsive case (top panels), the soliton is primarily
reflected by the defect (with a small transmitted fraction of the
norm). Similar behavior was observed for other values of
$\protect\zeta $, with the amount of material passing through the
defect increasing with $\protect\zeta _{0}$. On the other hand, in
the attractive case (bottom panels), the soliton gets fragmented
into reflected, trapped and transmitted parts. For larger initial
values of $\protect\zeta $, the trapped fraction is smaller.}
\label{Fig12}
\end{figure}

Having examined statics and dynamics in the vicinity of the stable and
unstable fixed points of the system, we now turn to an investigation of the
dynamics, setting the initial soliton farther away from the equilibrium
positions. Figure\ \ref{Fig12} displays three typical examples, with the
soliton set to the left and to the right of the repulsive (and of the
attractive) defect. In the repulsive case, we observe that the soliton is
primarily reflected from the defect; however, when it has large kinetic
energy at impact (which takes place if it was initially put at a position
with large potential energy), a fraction of the matter is transmitted
through the defect. On the other hand, in the attractive case, a fraction of
the matter is always trapped by the defect. However, this fraction is
smaller when the kinetic energy at impact is larger.

We note in passing that, while Eq.\ (\ref{meq8}) can predict not
only the equilibrium positions of the soliton but also the
dynamical behavior of $\zeta (t)$, we have opted not to use it for
the numerical experiments. The main reason is that, as can be
clearly inferred from Figs.\ \ref{Fig10}-\ref{Fig11} and
\ref{Fig12}, the interaction of the solitary wave with the defect
entails emission of a sizable fraction of matter in the form of
small-amplitude waves, which, in turn, may interfere with the
solitary wave and significantly alter his motion (see e.g., Fig.\
\ref{Fig12}). Hence, the prediction of the dynamics based on the
adiabatic approximation, which is implied in Eq. (\ref{meq8}),
would be inadequate in the presence of these phenomena.

\section{Conclusions\label{sec:conc}}

In this work, we have examined the interaction of bright solitary waves with
localized defects in the presence of magnetic trapping, which is relevant to
Bose-Einstein condensates with negative scattering lengths. We have found
that the defect induces, if its strength is sufficiently large, the
existence of two additional steady states (bifurcating into existence
through a saddle-node bifurcation), one of which is stable and one unstable.
We have constructed the relevant bifurcation diagram and explicitly found
both the stable and the unstable solutions, and quantified the instability
of the latter via the presence of a real eigenvalue pair. The dynamical
instability of these unstable states leads to oscillations around (for
repulsive defects) and/or trapping at (for attractive defects) the nearby
stable steady state. Additionally, we have developed a collective-coordinate
approximation to explain the steady soliton solutions and the 
corresponding bifurcation.
We have illustrated the numerical accuracy of the analytical
approximation by comparison with direct numerical results. We have also
displayed, through direct numerical simulations, the dynamics which follows
setting the initial soliton off a steady-state location. Noteworthy
phenomena that occur in this case are the emission of radiation by the
soliton colliding with the repulsive defect, and capture of a part of the
matter by the attractive one.

These results may be relevant to the trapping, manipulation and guidance of
solitary waves in the context of BEC. They illustrate the potential of the
combined effect of magnetic and optical (provided by a focused laser beam)
trapping to capture (either at or near the laser-beam-induced local defect)
a solitary wave which can be subsequently guided, essentially at will.
Naturally, the beam's intensity must exceed a critical value, which can be
explicitly calculated in the framework of the developed theory. It would be
particularly interesting to examine the predicted soliton dynamics in BEC
experiments.

\onecolumn

\end{document}